\documentclass{aastex}
\usepackage[onecolumn]{emulateapj5}

\def \Halpha{{H$\alpha$\ }}
\def \eg{{e.g.,}}

\def \etal{{et~al.\null}}

\shorttitle{Planetary Nebulae as Standard Candles}

\shortauthors{Ciardullo et al.}

\begin{document}

\title{Planetary Nebulae As Standard Candles.~XII.  \\
Connecting the Population~I and Population~II Distance Scales}

\author{Robin Ciardullo\altaffilmark{1}}
\affil{Department of Astronomy \& Astrophysics, The Pennsylvania State
University
\\ 525 Davey Lab, University Park, PA 16802}
\email{rbc@astro.psu.edu}

\author{John J. Feldmeier\altaffilmark{1}}
\affil{Department of Astronomy, Case Western Reserve University \\10900 Euclid
Ave., Cleveland, OH 44106-1712}
\email{johnf@eor.cwru.edu}

\author{George H. Jacoby}
\affil{WIYN Observatory, P.O. Box 26732, Tucson, AZ 85726}

\email{jacoby@wiyn.org}

\and

\author{Rachel Kuzio de Naray, Mary Beth Laychak, Patrick R. Durrell}
\affil{Department of Astronomy \& Astrophysics, The Pennsylvania State
University
\\ 525 Davey Lab, University Park, PA 16802}
\email{kuzio@astro.umd.edu, laychak@astro.psu.edu, pdurrell@astro.psu.edu}

\altaffiltext{1} {Visiting Astronomer, Kitt Peak National Observatory,
National Optical Astronomy Observatory, which is operated by the Association
of Universities for Research in Astronomy, Inc. (AURA) under cooperative
agreement with the National Science Foundation.}

\begin{abstract}
We report the results of [O~III] $\lambda 5007$ surveys for planetary
nebulae (PNe) in six galaxies: NGC~2403, NGC~3115, NGC~3351, NGC~3627,
NGC~4258, and NGC~5866.  Using on-band/off-band [O~III] $\lambda 5007$ images,
as well as images taken in H$\alpha$, we identify samples of PNe in these
galaxies and derive their distances using the planetary nebula luminosity
function (PNLF).  We then combine these measurements with previous data to
compare the PNLF, Cepheid, and surface brightness fluctuation (SBF) distance
scales.  We use a sample of 13 galaxies to show that the absolute magnitude
of the PNLF cutoff is fainter in small, low-metallicity systems, but the
trend is well modeled by the theoretical relation of \citet{djv92}.  When this
metallicity dependence is removed, the scatter between the Cepheid and
PNLF distances becomes consistent with the internal errors of the methods and
independent of any obvious galaxy parameter.  We then use these data to
recalibrate the zero point of the PNLF distance scale.  We use a sample
of 28 galaxies to show that the scatter between the PNLF and SBF distance
measurements agrees with that predicted from the techniques' internal errors,
and that there is no systematic trend between the distance residuals and
stellar population.  However, we also find that the PNLF and SBF methods have
a significant scale offset:  Cepheid-calibrated PNLF distances are, on
average, $\sim 0.3$~mag smaller than Cepheid-calibrated SBF distances.  We
discuss the possible causes of this offset, and suggest that internal
extinction in the bulges of the SBF calibration galaxies is the principle cause
of the discrepancy.   If this hypothesis is correct, then the SBF-based Hubble
Constant must be increased by $\sim 7\%$.  We also use our distance to
NGC~4258 to argue that the short distance scale to the LMC is correct, and
and that the global Hubble Constant inferred from the {\sl HST\/} Key Project
should be increased by $8 \pm 3\%$ to $H_0 =  78 \pm 7$~km~s$^{-1}$~Mpc$^{-1}$.

\end{abstract}


\section{Introduction}
The past two decades has seen remarkable progress in the measurement of the
distance scale of the universe.  In the early and mid-1980's, values of the
Hubble Constant ranged over a factor of two, from
$H_0 \approx 50$~km~s$^{-1}$~Mpc$^{-1}$ \citep[\eg][]{kkct88, st82} to
$H_0 \approx 100$~km~s$^{-1}$~Mpc$^{-1}$ \citep[\eg][]{deV85, huc87}, with
the results depending strongly on the the author and the technique.
However, in the early 1990's, measurements of $H_0$ began to converge
\citep{mudville}, and it became increasingly difficult to argue for values
much different than $H_0 \sim 70$~km~s$^{-1}$~Mpc$^{-1}$.  Today, due in
large part to the {\sl Hubble Space Telescope\/} Distance Scale Key Project,
a value of $H_0$ between 65 and 75 km~s$^{-1}$~Mpc$^{-1}$ is generally
accepted \citep{keyfinal}.

Nevertheless, there are still two lingering problems with the current distance
scale.  The first concerns the zero point of the Cepheid period-luminosity
relation.  There are two galaxies whose distances are known from direct
geometric techniques: the Large Magellanic Cloud \citep[via the light echo of
SN~1987A;][]{panagia91, gould}, and NGC~4258 \citep[through the observed
motions of its nuclear maser;][]{herrnstein}.  The former sets the zero point
for the Cepheid scale; the latter provides an independent test of the
technique.  Unfortunately, the Cepheid distance to NGC~4258 given by
\citet{keyfinal} is $1.2 \sigma$ larger than the galaxy's geometric distance
\citep{herrnstein}.  This marginally significant discrepancy may indicate
a problem for the zero point of the system.

The second limitation of the Cepheid distance scale is its limited
applicability to Population~II distance techniques.  For example, the
calibration of the elliptical galaxy fundamental plane \citep{kelson, keyfinal}
rests largely on the assumption that the early-type galaxies of Leo~I, Fornax,
and Virgo are at the same distance as the clusters' spirals.  In the case of
Virgo at least, this is likely not the case \citep{cm87, wb00}.  Similarly,
the zero point of the surface brightness fluctuation (SBF) technique
\citep{ferrarese, tonry, keyfinal} is set by just six Cepheid calibrators.
Given the susceptibility of the SBF method to the effects of interstellar
extinction, this situation is not ideal.  Clearly, additional calibrators
are needed to secure the Pop~II side of the distance ladder.

The planetary nebula luminosity function (PNLF) has the potential to provide
these calibrations.  As the only general purpose standard candle that is
applicable to both spiral and elliptical galaxies, the PNLF provides a
critical link between the Pop~I and Pop~II distance scales.   Moreover,
since the precision of the PNLF method is comparable to that of Cepheids
\citep{mudville}, the technique can also be used to check for anomalous
measurements in the distance ladder.  In fact at present, the PNLF is the
only method capable of confirming the results obtained from Cepheid variables.
Finally, the PNLF can provide distances to some intermediate objects that
are too dusty or irregular for Pop~II techniques, but not suitable for
Cepheid observations.

In this paper, we present the PNLFs of six galaxies, NGC~2403, 3115, 3351,
3627, 4258, and 5866, and use these data to search for systematic errors in the
extragalactic distance ladder.  In Section 2, we describe our observations,
detail our reduction procedures, and present the coordinates and [O~III]
$\lambda 5007$ magnitudes of our planetary nebula candidates.  We also present
new [O~III] and \Halpha observations of PNe in the inner bulge of M31;
these data are used in Section~3 to create a quantitative criterion
for discriminating PNe from compact H~II regions.  In Section~4, we derive
PNLF distances to our six galaxies, and comment on the properties of these
systems.  Included in this section is a discussion of the distance to
NGC~4258; our value, combined with that obtained from the Cepheids, argues
for a Hubble Constant that is $\sim 7\%$ larger than that given by the
{\sl HST\/} Key Project \citep{keyfinal}.  In Section~5, we combine our
distances to NGC~2403, 3351, 3627, and 4258 with data from nine other
Cepheid galaxies to re-define the zero point of PNLF distance scale.
We show that the absolute magnitude of the PNLF bright-end cutoff does shift
to fainter magnitudes at extremely low metallicity;  this is in agreement
agreement with the theoretical predictions of \citet{djv92}.  However, we
show that in metal-rich galaxies, the PNLF-Cepheid residuals show no
statistically significant trend.  In Section~6, we compare the PNLF distance
scale with that of the SBF method, and show that there is a significant scale
error between the two techniques.  Specifically, we show that, although the
PNLF-SBF residuals do not correlate with any galaxy property, the overall PNLF
scale is $\sim 0.3$~mag shorter than the SBF scale.  Finally, we conclude by
considering the possible causes of this discrepancy, and discussing the
implications it has for tip of the red giant branch (TRGB) distance
measurements and the extragalactic distance scale in general.

\section{Observations and Reductions}
Several different telescope/detector configurations were used in this program.
Four galaxies were observed with the Kitt Peak 4-m telescope:  NGC~3115,
with a $320 \times 512$ RCA CCD (43~e$^-$ readnoise), NGC~5866 with an
$800 \times 800$ TI CCD (binned $2 \times 2$, 4~e$^-$ readnoise), and
NGC~2403 and NGC~3627 with the T2KB $2048 \times 2048$~CCD (4~e$^-$ readnoise).
Two other galaxies, NGC~3351 and NGC~4258, were observed with the Mini-Mosaic
camera of WIYN; this instrument employs two 4K $\times$ 2K CCDs covering a
field-of-view of $9\farcm 6 \times 9\farcm 6$, with a $7\farcs 1$ gap between
detectors.  Finally, in order to define a quantitative criterion for
discriminating planetary nebulae from compact H~II regions, the inner bulge
of M31 was observed with the Kitt Peak 2.1-m telescope and a Tektronix
$1024 \times 1024$ detector.  This setup allowed us to survey the central
$2\farcm 6$ region of the galaxy and measure monochromatic fluxes for
PNe $\sim 4$~mag down the [O~III] $\lambda 5007$ luminosity function.
A log of our observations and the details of the CCDs are given in Table~1.

Our survey technique was similar to that described in previous papers of the
series \citep{p3, p4, p11}.  First, a narrow-band filter was selected which
passed the [O~III] $\lambda 5007$ emission line at the redshift of the target
galaxy.  For M31, 2403, and 4258, the filters used were the standard
narrow-band [O~III] filters available at NOAO;  for the remaining galaxies,
custom-made filters were used.  The filter names or bandpasses (central
wavelength and full-width-half-maximum (FWHM) at ambient temperature
in the converging beam of the telescope) are given in Table~1.   Note that
for NGC~3351, our filter vignetted roughly half of the WIYN camera's
field-of-view.  This did not compromise our survey, however, since the
unvignetted field-of-view ($4\farcm 8 \times 4\farcm 8$) was reasonably well
matched to the $\sim 7\farcm 4 \times 5\farcm 0$ diameter of the galaxy.

For each galaxy, a series of exposures was taken through the on-band
[O~III] $\lambda 5007$ and a wider off-band filter ($\lambda_c = 5300$,
FWHM $\sim 250$~\AA).  In addition, in order to discriminate PNe from
H~II regions, the late-type spirals were also imaged through
a 75~\AA\ wide filter centered on H$\alpha$.  The total on-band and
\Halpha exposure times are listed in Table 1.  In general, the off-band
exposures were scaled to go $\sim 0.2$~mag deeper than their on-band
counterparts and were therefore four to seven times shorter.  All frames were
reduced using standard IRAF routines:  the images were bias-subtracted and
flat-fielded using CCDPROC, aligned to a common astrometric system with
GEOMAP and GEOTRAN, and then combined using IMCOMBINE.  The result was
a set of summed on-band, off-band, and \Halpha images of each galaxy.

Planetary nebulae were identified on our frames in two complementary ways:
by ``blinking'' the summed [O~III] image against the off-band image, and by
looking for emission-line objects on a ``difference'' image, formed by
subtracting a scaled off-band image from the on-band frame.  In order to be
considered a planetary nebula candidate, an object had to have a
point-spread-function consistent with that of a point source, and be
present on the on-band image, but completely invisible on the off-band frame.
PN candidates in star-forming galaxies also had to be at least 1.6 times
brighter in [O~III] than in \Halpha (see Section 3).

The PN candidates were measured photometrically using the IRAF version of
DAOPHOT \citep{stet87, stet90, stet92}, and flux calibrated using
large-aperture measurements of \citet{stone} and \citet{massey} standard stars
and the procedures outlined by \citet{jqa}.  Our 4-m and WIYN observations
were taken under photometric conditions, so this straightforward calibration
produced fluxes with a zero-point accurate to $\sim 0.04$~mag.  The M31
data, however, were obtained through light cirrus.  The PNe in this galaxy
were therefore calibrated by scaling the data to the [O~III] $\lambda 5007$
and \Halpha observations of M31's bulge taken by \citet{p2} and \citet{cfnjs}.
Once calibrated, the resulting [O~III] monochromatic fluxes (in
ergs~cm$^{-2}$~s$^{-1}$) were converted to $m_{5007}$ magnitudes using
\begin{equation}
m_{5007} = -2.5 \log F_{5007} - 13.74
\end{equation}

Tables~2 through 8 list the PNe of each galaxy.  For M31, the equatorial
positions are based on the J2000 system of the Guide Star Catalog \citep{gsc}
and a set of secondary standards defined by \citet{hui94}.  For the remaining
galaxies, the plate solutions were created using the USNO-A2.0 catalog
\citep{monet}.  The errors associated with our positions are $\sim 0\farcs 4$.
Note that the PN identifications in Table~2 are an extension of the numbering
scheme of \citet{p2}.  Figure~1 displays our [O~III] $\lambda 5007$ images of
the six galaxies, with the positions of the PN candidates marked with crosses.

\section{Discriminating Planetary Nebulae from H~II Regions}
PNLF measurements in elliptical and S0 galaxies, such as NGC~3115 and NGC~5866,
are relatively straightforward: since these systems have no current star
formation, virtually all bright emission-line sources are planetary nebulae.
PN observations in late-type galaxies, however, are not so simple.  Only a
very small percentage of the emission-lines sources in these systems are
planetary nebulae; most [O~III] line-emission comes from H~II regions and
supernova remnants.  Therefore, in order to derive PN distances to these
star-forming systems, an algorithm is needed to discriminate PNe from other
sources of line emission.

As stated above, we used three criteria to perform this discrimination.
In order to be classified as a planetary nebula, an object had to be spatially
unresolved, invisible on the off-band frame, and have an [O~III] to \Halpha
ratio greater than 1.6.  The first of these conditions is obvious.  All bright
PNe in the Galaxy are less than $\sim 1$~pc across \citep{acker};  in M31,
this corresponds to an angular size of $\sim 0\farcs 27$.  The true
PNe in our program galaxies must therefore be stellar; objects that are
even marginally resolved must be either H~II regions or supernova remnants.

Our second criterion is also quite simple.  Although the central stars
of PNe can have luminosities $\log L/L_\odot \gtrsim 4$, their effective
temperatures are such that very little of this energy comes out in the
optical.  At the distances considered here, central star continuum emission at
5300~\AA\ is well below the threshold of detectability.  On the other hand,
H~II regions usually have OB associations at their center.  Depending on how
many stars are present, the OB stars' combined optical emission may be
detectable in the continuum.  Thus, all valid PN candidates must be completely
invisible on the off-band frame.

The third criterion, that of $R = I(\lambda 5007) / I({\rm H}\alpha +
{\rm N~II}) > 1.6$ requires some elaboration.  The central stars of [O~III]
bright PNe are typically much hotter than the OB stars that excite H~II
regions.  As a result, much of the oxygen in a planetary nebula is doubly
ionized (rather than singly ionized as is common in most H~II regions).  This
fact, combined with the higher electron temperatures produced by the harder
radiation field, causes a typical bright PN to have [O~III] $\lambda 5007$ much
brighter than H$\alpha$.  Conversely, as the survey of \citet{shaver}
illustrates, most H~II regions ($\gtrsim 80\%$) have H$\alpha$+[N~II] brighter
than [O~III] $\lambda 5007$.

We can be quantitative about this condition by defining $R$ as the [O~III] to
H$\alpha$+[N~II] line ratio and examining the behavior of $R$ as a function of
[O~III] $\lambda 5007$ absolute magnitude.  [O~III] $\lambda 5007$, H$\alpha$,
and [N~II] line strengths now exist for large samples of PNe in three Local
Group galaxies:  M33, through the [O~III] and H$\alpha$+[N~II] photometry of
\citet{magrini33a, magrini33b}, the Large Magellanic Cloud, through the [O~III]
photometry of \citet{p6} and the spectrophotometric line ratios of \citet{md1},
\citet{md2}, and \citet{vdm}, and M31 (this paper).  The data are plotted in
Figure~2.  As the figure shows, the distribution of line ratios for [O~III]
bright planetary nebulae is remarkably independent of stellar population.
PNe that are several magnitudes down the [O~III] $\lambda 5007$ luminosity
function have values of $R$ that are anywhere from $\sim 1/4$ to $\sim 4$.
Objects in the top $\sim 1$~mag of the PNLF, however, all have $R \gtrsim 2$.
This result holds for all populations, from the old, metal-rich bulge stars
of M31, to the young, metal-poor planetaries of the LMC{}.  Conversely, H~II
regions with $R \gtrsim 2$ are rare: of the 42 objects studied by
\citet{shaver}, only four have this ratio greater than 1.5, and only one has
$R > 2.2$.

Since the [O~III] $\lambda 5007$ surveys presented in this paper extend less
than $\sim 1$~mag down the PNLF, an efficient criterion for the removal of
contaminating H~II regions is to consider only those objects with [O~III]
$\lambda 5007$ to H$\alpha$+[N~II] line ratios greater than 2.  Note that this
is a more stringent requirement than the $R > 1$ criterion imposed by
\citet{magrini33a} and \citet{magrini81} in their PN surveys of M33 and M81.
However, Magrini \etal\ derived their value by considering the excitations of
all PNe,  regardless of absolute [O~III] $\lambda 5007$ magnitude.  As
Figure~2 demonstrates, a criterion that uses absolute magnitude in addition
to excitation is a much more powerful discriminator.

Unfortunately, due to time constraints and variable seeing at the telescope,
our \Halpha images did not go deep enough to record objects with $R \sim 2$.
Instead, for the faintest PNe in our sample, our \Halpha limiting magnitude
corresponded to $R = 1.6$.  We therefore used this number as our PN/H~II
region discriminator.   This criterion is not perfect:  if just 1\% of
bright H~II regions are high excitation objects, then a substantial number
of interlopers may survive this cut.  However, by using our excitation rule
in combination with the other two conditions, and limiting our PN search
to regions away from the galaxies' spiral arms and obvious star-forming
regions, we are confident that we have reduced the fraction of contaminants
to a negligible level.

\section{Fitting the PNLFs and Obtaining Distances}
Figure~3 displays the [O~III] $\lambda 5007$ PNLFs for the six galaxies in our
sample.  Each shows an abrupt rise, and a flattening that is characteristic
of the PNLF of other galaxies \citep[see][]{mudville}.  However, before
these data can be used to derive distances, we need to define statistically
complete samples for analysis.

For NGC~3115 and NGC~5866, this task was relatively straightforward:
we used the fact that, on smooth backgrounds, a $\sim 100\%$ detection
rate occurs at a signal-to-noise of $\gtrsim 10$ \citep{cfnjs, hui93}.
The PN limiting magnitude is therefore a unique function of the background
surface brightness: by noting the background at each position in the
galaxies, we could define photometrically complete samples of objects
\citep[see][]{p3, p4}.  Unfortunately, in the spiral galaxies, variable
internal extinction, as well as the complexity of the underlying background
makes this type of analysis impossible.  Thus, statistical sub-samples of PNe
had to be derived empirically from the frames.

To create the PN samples for the later-type galaxies, we began by noting
the median sky background associated with each PN measurement.   After
excluding those few objects superposed on bright regions of the galaxy,
we picked the worst (most uncertain) background remaining in the sample,
and computed the signal-to-noise each PN would have had, if it had been
projected on that background.  We then defined our statistical PN sample
as those objects with a hypothetical signal-to-noise above some threshold
value.  In the case of NGC~3351, 3627, and 4258, a threshold signal-to-noise
of 10 served to define our limiting PN magnitude.  For NGC~2403, however, our
ability to blink was restricted by the complexity of the galaxy's emission
regions.  Simply put, our PN identifications in NGC~2403 were confusion
limited, rather than detection limited.  As a result, we chose an extremely
high value for the detection threshold for this galaxy -- only objects with a
hypothetical signal-to-noise greater than 25 made it into our complete sample.
Those PNe that are part of our statistical samples are identified in Tables 3-8
with an ``S''.

In order to derive PNLF distances and their formal uncertainties, we followed
the procedure of \citet{p2}.  We took the analytical form of the PNLF
\begin{equation}
N(M) \propto e^{0.307 M} \{ 1 - e^{3 (M^* - M)} \}
\end{equation}
convolved it with the photometric error vs.~magnitude relation derived from
the DAOPHOT output, and fit the resultant curve to the statistical
samples of PNe via the method of maximum likelihood.   To correct for
foreground extinction, we used the $100 \mu$ DIRBE/IRAS all-sky map of
\citet{schlegel}, and the reddening curve of \citet{ccm}.  Finally, to
estimate the total uncertainties in our measurements, we convolved the formal
errors of the maximum-likelihood fits with the errors associated with the
photometric zero points of the CCD frames (0.04~mag), the filter response
curves (0.03~mag), and the Galactic foreground extinction
\citep[$0.16 \, E(B-V)$;][]{schlegel}.  For consistency with previous papers
of the series, we used a PNLF zero point of $M^* = -4.48$, which is based
on an M31 distance of 710~kpc \citep{welch} and foreground reddening of
$E(B-V)=0.11$ \citep{mcclure}.  We will revisit the question of the PNLF zero
point in Section~5.

\subsection{NGC 2403}
NGC~2403, the medium-sized Scd spiral in the M81 Group, has historically
been the limit for ground-based Cepheid observations, and its
Cepheid measurements date back over half a century \citep{ts68}.  The
galaxy's revised Cepheid distance modulus of $(m-M)_0 = 27.48 \pm 0.10$
\citep{keyfinal} is based on the $I$-band photometry
of 10 Cepheids with the Canada-France-Hawaii telescope \citep{fm88}
and the assumption that the galaxy's internal extinction is similar to
that of other spirals.

At a distance of $\sim 3$~Mpc, it is extremely easy to detect NGC~2403's
planetaries and resolve out the galaxy's H~II regions, even in $1\farcs 3$
seeing.  Consequently, most of our PNe are extremely well detected, with
signal-to-noise values over 100.  Nevertheless, the PNLF of NGC~2403 turns
over just $\sim 1$~mag down the luminosity function, as if the data past
$m_{5007} \sim 24$ were severely incomplete.  Part of the reason for this
behavior may be due to the fact that in many regions of the galaxy, PN
detections were confusion limited.  However, this peculiar feature of the
luminosity function may also be intrinsic to the galaxy's stellar population.
\citet{jd02} have shown that the luminosity function of PNe in the SMC is
non-monotonic with a pronounced dip $\sim 3$~mag down the luminosity function.
The luminosity function of \citet{magrini33a} suggests that M33 has a
similar feature $\sim 1$~mag down the PNLF{}.  Like the SMC and M33,
NGC~2403 is undergoing strong star formation.  It is therefore possible that
the turnover in the luminosity function at $m_{5007} \sim 24$ real.  If
so, the position and strength of the dip may serve as an age or metallicity
indicator for the stellar population \citep{cks}.

Even if NGC~2403's PNLF declines past $m_{5007} \sim 24$, this behavior
does not affect the precision of the PNLF method, since PNLF distances
depend only on the brightest objects.  A fit to the top $\sim 1$~mag of the
observed PNLF yields a distance modulus for the galaxy
of $(m-M)_0 = 27.65_{-0.12}^{+0.07}$ ($D = 3.4_{-0.14}^{+0.11}$~Mpc).
This is $\sim 1\, \sigma$ larger than that derived from the Cepheids.

\subsection{NGC~3115}
Due to the small size of the RCA CCD, this prototypical field S0 galaxy
was observed in two parts, with the fields positioned $\sim 100\arcsec$
north and south of the galaxy's nucleus.  The inner regions of the
galaxy were excluded from our analysis, since PN detections in these high
surface brightness regions ($B < 22.2$~mag~arcsec$^{-2}$) are difficult.
Despite this limitation, we were still able to identify a statistical sample of
29~PNe that extends $\sim 1$~mag down the luminosity function.  If we fit these
data to the empirical curve, then the galaxy's most likely distance modulus
is $(m-M)_0 = 30.03_{-0.14}^{+0.11}$ ($D = 10.1^{+0.5}_{-0.6}$~Mpc).

Because NGC~3115 has a smooth luminosity profile with little, if any,
internal extinction, it is possible to normalize our PN counts to
the bolometric luminosity of the sampled population.  This quantity,
defined as $\alpha_{2.5}$ by \citet{p2}, is potentially useful, since
it has been observed to vary by almost an order of magnitude in different
stellar populations \citep{c95}.

To perform our normalization, we adopted the surface photometry measurements
of \citet{hamabe}, and summed the $B$-band flux contained in our survey
regions.  We then converted this total $B$ magnitude to a bolometric magnitude
by integrating the galaxy's spectral energy distribution, as defined by the
multicolor optical and IR photometry of \citet{deVLongo} and \citet{persson}.
We then used this bolometric magnitude to compute $\alpha_{2.5}$ using the
maximum-likelihood technique described in \citet{p2}.  The resulting value,
$\alpha_{2.5} = 25.7^{+9.9}_{-5.2} \times 10^{-9}$~PN~yr$^{-1} L_\odot^{-1}$,
is similar to that found for the bulge of M31, and is typical of
that for an old, metal-rich stellar population \citep{c95}.

\subsection{NGC 3351}
The importance of the Leo~I Group to extragalactic distance measurements has
been recognized for almost half a century \citep{hms}.  The group is compact,
well-defined, and contains both early and late-type galaxies; it is thus
the ideal location for linking the Pop~I and Pop~II distance scales.
As a medium-sized SBb galaxy in the core of Leo~I, NGC~3351 is especially
valuable.  The galaxy is late enough to contain a large number of Cepheids
\citep{graham97} and be useful as a Tully-Fisher calibrator \citep{macri},
yet early enough to have a bulge that can be analyzed using the surface
brightness fluctuation method \citep{jensen, wat01}.

Because our [O~III] $\lambda 5007$ filter vignetted part of the field, our
PN survey of NGC~3351 was limited to the inner $2\farcm 4$ of the galaxy.
Nevertheless, we detected 20~PNe; 12 of these were bright enough to be
part of a complete sample.  Our derived distance modulus of
$(m-M)_0 = 30.05^{+0.08}_{-0.16}$ ($D = 10.2^{+0.4}_{-0.7}$~Mpc) agrees with
the $(m-M)_0 = 29.85 \pm 0.09$ value derived from the {\sl HST\/} photometry of
49 Cepheids \citep{graham97, keyfinal}.  More importantly, as demonstrated
in Figure~4, our PNLF distance to NGC~3351 is also in excellent agreement with
the PNLF distances to four other members of the group.  This consistency
provides further evidence that the absolute magnitude of the PNLF does not
depend strongly on galaxy type or stellar population.

\subsection{NGC 3627}
NGC 3627, a large Sb spiral in the Leo Triplet, became important for distance
scale research in 1989, when it produced the well-observed Type~Ia supernova
1989B \citep{wells}.  The galaxy has also hosted two other supernova-like
events, the Type II SN 1973R \citep{ciatti} and the unusual object SN 1997bs
\citep{vandyk}, and is a zero-point calibrator for the Tully-Fisher relation.

The body of NGC 3627 is quite dusty, so PN identifications in the bulge
and interarm region of the galaxy are difficult.  However, NGC~3627 has
recently undergone an interaction with NGC~3628; this is evidenced by the
galaxy's asymmetric spiral arms, the plumes of extragalactic H~I and optical
material extending from NGC~3628 \citep{rots, kb74, chromey}, and the large
faint halo surrounding NGC~3627 \citep{burk81}.  It is, in fact, this low
surface brightness stellar halo that allows us to determine the galaxy's
distance; the region contains many bright planetary nebulae and very few H~II
regions.

Our survey of NGC~3627's halo produced 73 planetary nebula candidates; 40 of
these constituted our statistically complete sample.  A maximum likelihood fit
to these data yields a distance modulus of $(m-M)_0 = 29.99^{+0.07}_{-0.08}$
($D = 10.0^{+0.3}_{-0.4}$~Mpc), in excellent agreement with the value of
$(m-M)_0 = 29.86 \pm 0.08$ obtained from {\sl HST\/} observations of 68
Cepheids \citep{saha99, keyfinal}.

\subsection{NGC 4258}
In the mid-1990's, nuclear masers were discovered moving in a Keplerian orbit
around the central black hole of NGC~4258 \citep{ww94}.  By resolving these
masers \citep{miyoshi}, measuring their acceleration \citep{greenhill},
and monitoring their proper motions, \citet{herrnstein} were able to derive
a geometric distance to the galaxy, $7.2 \pm 0.3$~Mpc.   This makes NGC~4258
one of only two galaxies with a direct geometric distance, and an important
crosscheck for extragalactic distance techniques.

Our [O~III] $\lambda 5007$ survey in NGC~4258 resulted in the identification
of 58~PN candidates.  Of these, 29 were above our signal-to-noise cutoff
of 10, and had magnitudes in the top $\sim 0.7$~mag of the PNLF{}.  The
distance implied by these PNe, $(m-M) = 29.42^{+0.07}_{-0.10}$
($7.6^{+0.2}_{-0.3}$~Mpc), is $\sim 1 \sigma$ larger than the system's
geometric distance.

The fact that our PNLF distance is slightly higher than the galaxy's
geometric distance may be due to chance.  However, it also may be indicative
of a more substantial problem with the extragalactic distance scale.
{\sl HST\/} measurements of 15 Cepheids in the disk of NGC~4258 yield a
distance modulus of $(m-M) = 29.44 \pm 0.12$ (random + systematic error)
or $7.7 \pm 0.4$~Mpc \citep{newman01, keyfinal}.  Like the PNLF distance,
the Cepheid value is also slightly high (by $1.2 \sigma$).

Is this a serious problem?  Aside from NGC~4258, there is only one other
galaxy with a geometrical distance estimate.  That galaxy, the Large
Magellanic Cloud, has a measurement based on the light echo of
SN~1987A \citep{panagia91}.  Unfortunately, the light-echo distance is
still somewhat controversial, with recent estimates differing by
$\sim 0.2$~mag, from $52.0 \pm 1.3$~kpc \citep{panagia99} to less than
$47.2 \pm 0.1$~kpc \citep{gould}.  The former distance is close to that
derived from the Galactic calibration of Cepheids \citep{fc97} and the
Hipparcos calibration of RR Lyrae variables \citep{reid}; the latter value
is in line with distances inferred from the statistical parallax
calibration of RR Lyrae stars \citep{layden} and the Hipparcos calibration
of red clump stars \citep{udalski00}.

The Cepheid distance scale of \citet{keyfinal} is based on an LMC distance
modulus of $(m-M)_0 = 18.50$.  However, the most comprehensive analysis of
SN 1987's light-echo is probably that of \citet{gould}, and their distance
modulus is $(m-M)_0 < 18.37 \pm 0.04$ (where the upper limit reflects the
possible time lag between the arrival of the EUV-light and line fluorescence).
If we adopt this shorter value, then the geometric distance ratio between the
LMC and NGC~4258 becomes $\Delta\mu_{\rm geom} = 10.92 \pm 0.10$.  This is in
perfect agreement with the distance ratios derived from both the Cepheids
($\Delta\mu_{\rm Cep} = 10.94 \pm 0.12$), {\it and\/} the PNLF
($\Delta\mu_{\rm PNLF} = 11.00 \pm 0.14$).

Our PNLF distance to NGC~4258, when combined with that from the Cepheids,
provides independent evidence in support of the ``short'' distance to the
LMC{}.  {\it If this result holds, then the entire extragalactic distance scale
is affected.}  Specifically, if we work backward from the \citet{herrnstein}
distance for NGC~4258, and apply the observed PNLF and Cepheid LMC/NGC~4258
distance ratios, then the LMC distance modulus becomes $18.33 \pm 0.13$.
This increases the \citet{keyfinal} Hubble Constant by 8\%, from
$H_0 = 72 \pm 8$~km~s$^{-1}$~Mpc$^{-1}$ to
$H_0 = 78 \pm 8$~km~s$^{-1}$~Mpc$^{-1}$.   Moreover, if we further
constrain the LMC distance using the \citet{gould} measurement of
SN~1987A's light echo, then the error on the LMC distance is reduced to
$\pm 0.07$~mag, and the \citet{keyfinal} Hubble Constant
becomes $H_0 = 78 \pm 7$~km~s$^{-1}$~Mpc$^{-1}$.

\subsection{NGC 5866}
The edge-on S0 galaxy NGC~5866 is the most distant galaxy in our sample, and
it has the poorest data quality.  As a result, PN identification could only be
made in the galaxy's halo, where the galactic surface brightness is well below
that of the sky.  Even then, our limiting magnitude for completeness reached
only $\sim 0.5$~mag down the PN luminosity function.  Still, the 11~PNe
in our statistical sample yield a relatively well-constrained distance
modulus of $(m-M)_0 = 30.75^{+0.08}_{-0.12}$ ($D = 14.1^{+0.5}_{-0.7}$~Mpc).

For galaxies such as NGC~5866 with distances greater than $\gtrsim 13$~Mpc,
PN samples can suffer contamination due to the redshifted emission lines
of distant galaxies.  It is therefore reasonable to consider the possibility
that some of the PN candidates listed in Table~8 are, in reality, Ly$\alpha$
galaxies at $z=3.13$.  In the case of NGC~5866, however, this source of
contamination should be negligible.  \citet{cfkjg} have shown that the surface
density of emission-line contaminants brighter than our limiting magnitude of
$m_{5007} = 26.7$ is $\sim 1$~object per 66~arcmin$^{-2}$.  Since NGC~5866
was surveyed with a small chip (16~arcmin$^{-2}$), the likelihood of finding
a background galaxy in the field is low.  Thus our PNLF distance should not be
affected by this source of error.

\section{The PNLF-Cepheid Comparison}
The PNLF is a secondary standard candle.  Although there have been a few
attempts to measure the absolute magnitude of the PNLF cutoff in the
Milky Way \citep[\eg][]{pottasch90, mendez93}, $M^*$ must still be
defined via observations in galaxies with known distances.  The original
value of the zero point, $M^* = -4.48$, was based on an M31 Cepheid distance
of 710~kpc \citep{welch} and a foreground extinction of $E(B-V)=0.11$
\citep{mcclure}.  Since then the Cepheid distance to M31 has increased
\citep[750~kpc;][]{keyfinal} and estimates of the foreground extinction
have decreased \citep[$E(B-V)=0.062$;][]{bh84, schlegel}.  Formally, this
brightens $M^*$ to $-4.53$ and increases all the PNLF distances by 2.5\%.

Rather than rely on a single galaxy for the PNLF zero point, it is better
to derive $M^*$ by combining the PNLF measurements of many different systems.
With the addition of NGC~2403, NGC~3351, NGC~3627, and NGC~4258, there are
now 13 galaxies with both Cepheid and PNLF photometry.  If we adopt the
final {\sl HST Key Project\/} Cepheid distances (uncorrected for
metallicity) given by \citet{keyfinal} and use the DIRBE/IRAS estimates for
foreground extinction \citep{schlegel}, we can obtain a measure of $M^*$ in
each system.  These are listed in Table~9 and plotted as a function of
galactic metallicity \citep[as estimated from the systems' H~II regions;][]
{fdatabase} in Figure~5.  Note that $M^*$ is not measured with the
same precision in each galaxy.  This is a limitation intrinsic to the PNLF
method: low-luminosity systems simply do not have as many PNe populating the
bright-end of the luminosity function as large, luminous galaxies \citep{p2}.

As Figure~5 illustrates, at low-metallicity, our derived values of $M^*$
appear to be sensitive to oxygen abundance.  In particular, in the three most
metal-poor galaxies in the sample (SMC, NGC~5253, and NGC~300) the PNLF zero
point is substantially ($0.19^{+0.04}_{-0.16}$) fainter than that seen in
the metal-rich systems.  This behavior is not unexpected: a decrease in the
[O~III] $\lambda 5007$ emission of metal-poor planetary nebulae has been seen
by \citet{p8} and \citet{richer} in Local Group surveys and the trend has been
successfully modeled by \citet{djv92}.  According to the \citet{djv92}
models, $M^*$ is brightest when the PN oxygen abundance is near solar;
in more metal-poor and metal-rich populations, $M^*$ fades.  This dependence,
which is plotted in Figure~5, can be expressed via the quadratic
\begin{equation}
\Delta M^* = 0.928 {\rm [O/H]}^2 + 0.225 {\rm [O/H]} + 0.014
\end{equation}
where the solar abundance of oxygen is assumed to be $12+\log {\rm (O/H)}
= 8.87$ \citep{grevesse}.  In reality, the weakening of [O~III] $\lambda 5007$
in high-Z galaxies will not be observed, since in these systems,
solar-metallicity stars are available to define the systems' PNLF cutoff.
However, the dimming of $M^*$ in low metallicity galaxies can be observed,
and, as Figure~5 indicates, the observed trend is in excellent agreement
with theory.

To search for further systematic errors in the PNLF, we applied the
\citet{djv92} metallicity correction to the galaxies on the low-metallicity
side of the relation's inflection point, and replotted the data against
several galactic parameters.  The results, displayed in Figure~6, are
impressive.  Except for one object (M33), the scatter in the data is perfectly
consistent with the internal uncertainties associated with the PNLF and
Cepheid distance measurements.  This implies that the quoted errors of the
two techniques are reasonable, and that any additional sources of error are
small.  Moreover, the one discrepant point may have an explanation.
The $\sim 0.35$~mag ($2.3 \, \sigma$) offset between M33's expected and
observed PNLF zero point is essentially the same as the offset found between
the galaxy's Cepheid distance and its distance derived from RGB tip and
red clump stars \citep{kim}.  In both cases, the discrepancy can be traced
to a difference in the adopted reddening.  M33's Cepheid distance uses an
extinction that is derived from multicolor photometry of the Cepheids
themselves \citep{freed91, keyfinal}.  Our measurement of $M^*$, and the
measurement of the galaxy's RGB stars, use the DIRBE/IRAS Galactic extinction.
The difference between these values, $E(B-V) = 0.17$, is one of the largest
observed for any Cepheid galaxy, and may be responsible for the anomalous
value of $M^*$.   If the Cepheid reddening were reduced to one typical of
other Cepheid galaxies, M33's PNLF cutoff (and the red star distance
indicators) would be in much better agreement.

Further evidence of the consistency of the PNLF technique comes from the lack
of correlation between the metallicity-corrected value of the PNLF zero point,
$M^*_Z = M^* + \Delta M^*(Z)$, and the various galaxy properties.  For
example, if the assumed shape of the PNLF (equation~2) were incorrect, as
hypothesized by \citet{bottinelli}, we would expect to see a correlation
between $M^*_Z$ and galaxy absolute magnitude; such a correlation does not
exist.  Similarly, if compact H~II regions were a problem for PN surveys, then
PNLF measurements in distant galaxies would be preferentially affected, and
$M^*_Z$ would correlate with distance.  This trend is not seen, either.

The third panel of Figure~6 plots $M^*_Z$ against galaxy inclination for
the 11 spiral galaxies in our sample.  Internal extinction can affect PNLF
measurements, and indeed, the M33 point moves into perfect agreement with
the rest of the sample if we assume that the Cepheid reddening also applies
to the planetaries \citep{magrini33a}.  Nevertheless, the evidence of Figure~6
suggests that internal extinction is {\it not\/} a major source of error: if
it were, then we would expect $M^*_Z$ to correlate strongly with galaxy
inclination.  No correlation is seen, and this constancy supports the claim
of \citet{p11} that the relatively large scale-height of planetary nebulae
limits the effect of internal extinction to $\lesssim 0.1$~mag.

Perhaps the most interesting part of Figure~6 concerns the lack of any residual
correlation between the metallicity-corrected value for the PNLF zero point and
galaxy metallicity, as determined from the oxygen abundance of H~II regions
\citep{fdatabase}.  The dependence of Cepheid distances on galaxy metallicity
is controversial.  Theoretical and empirical analyses by numerous groups
disagree on both the amplitude and sign of the relation \citep[see][and
references therein]{feast, keyfinal}.  For example, while \citet{kochanek}
and \citet{sasselov} point out that metallicity corrections to the Cepheid
period-luminosity relation might change the extragalactic distance scale by
more than 10\%, other studies, such as those by \citet{caputo}, \citet{kenn98},
and \citet{udalski01} argue for a much weaker dependence.  Since our
measurements of $M^*_Z$ assume that the raw Cepheid distances (with no
metallicity correction) are accurate, the constancy seen in Figure~6 supports
the latter conclusion.

To quantify this result, we performed a Monte Carlo simulation on the data
displayed in the fourth panel of Figure~6.  We treated each galaxy's
asymmetrical error bars as probability distributions, and regressed $M^*_Z$
against metallicity for 1,000,000 realizations of the data.  The simulations
confirmed what is apparent to the eye:  once the PNLF zero point is corrected
for metallicity via the relation of \citet{djv92}, no other
term is needed.  Formally, the best-fitting slope to the relation is
$\Delta M^*_Z = 0.14 \pm 0.22$~mag~per~dex.  Either the metallicity
dependence of Cepheids exactly cancels that of an additional unknown PNLF
variation, or abundance changes have little effect on the distances derived
from Cepheids.

Since $M^*_Z$ does not appear to correlate with any galaxy property, we can
combine the probability distributions of Figure~6 and re-determine the
PNLF zero point.  Once again, we did this via a Monte
Carlo simulation in which we drew randomly from the probability distributions
associated with each galaxy.  If we exclude the outlying galaxy M33 from the
analysis, then our procedure yields a most probable value of $M^*_Z = -4.51$
with 68.2\% of the probability lying between $-4.49$ and $-4.55$.  (If M33
is included, this value drops slightly to $-4.48^{+0.02}_{-0.04}$.)
Alternatively, if we restrict our analysis to galaxies as large or larger
than the LMC ($12+\log {\rm O/H} \ge 8.5$), then the \citet{djv92} metallicity
correction is not needed and the most probable value of the zero point becomes
$M^* = -4.47^{+0.02}_{-0.03}$ (or $M^* = -4.43^{+0.02}_{-0.03}$ if M33 is
included in the sample).  Since emission-line abundances for most elliptical
galaxies are unobtainable, we will use $M^* = -4.47$ in the analysis that
follows and exclude low-luminosity galaxies from the analysis.

\section{The PNLF-SBF Comparison}
PNLF distances provide a bridge between the Pop~I distance indicators, such
as Cepheids, and the techniques on the Pop~II side of the distance ladder.
Chief among these Pop~II techniques is the Surface Brightness Fluctuation
method:  SBF distances have a precision comparable to that of Cepheids
and the PNLF, and the technique has been applied to over 300 elliptical,
lenticular, and early-type spiral galaxies \citep{tonry}.  Presently, the
SBF scale is calibrated via Cepheid measurements to six intermediate-type
spirals:  M31, M81, NGC~3368, NGC~4258, NGC~4725, and NGC~7331
\citep{keyfinal}.   Our PNLF measurements provide an independent test
of this calibration.

With the addition of NGC~3115, NGC~4258, and NGC~5866, there are now 28
large galaxies with both SBF and PNLF distance measurements.  These galaxies
are identified in Table~10, along with the differences between the
SBF and PNLF distance moduli.  For consistency with Section~5, and
to explicitly include the effects of cosmic scatter in the distance
indicators, the zero-point of the SBF method has been re-computed in
exactly the same manner as that for the PNLF{}.  That is, the SBF and
Cepheid error bars for the six calibrating galaxies have been added
in quadrature, and the resulting probability distributions used in a series
of Monte Carlo simulations.  This procedure produces an SBF distance scale
that is $0.04^{+0.04}_{-0.04}$~mag smaller than that used by
\citet{tonry}.

Figure~7 histograms the difference between the SBF distance moduli (as
calibrated directly via the six Cepheid galaxies) and the PNLF distance
moduli as calibrated in Section~5.  The curve displayed in the figure is the
expected scatter in the data, as determined by adding (in quadrature) the
uncertainties associated with the individual PNLF and SBF measurements, and
a $\sigma_{E(B-V)} = 0.16 E(B-V)$ uncertainty associated with Galactic
reddening \citep{schlegel}.  The latter component is especially important.  As
pointed out by \citet{cjt}, the strong color dependence of the absolute
fluctuation magnitude \citep[$\bar M_I \propto 4.5 (V-I)_0$;][]{tonry} means
that an underestimate of foreground extinction translates into an underestimate
of SBF distance.  This is the exact opposite of how the PNLF (and most other
methods) react to reddening:  for the PNLF, an underestimate of extinction
results in an overestimate of distance.  Consequently, a very small error
in extinction propagates into a significant discrepancy between the PNLF
and SBF moduli, with $\sigma_{\Delta \mu} = 7 \, \sigma_{E(B-V)}$.

The results of Table~10 and Figure~7 are noteworthy.  Immediately
obvious from the figure is the presence of three outliers.  NGC~4565,
NGC~891, and NGC~4278 have values of $\Delta\mu$ that are $\gtrsim 3\sigma$
from the mean of the distribution; this separation is too large to be
explained by any random process.  Interestingly, two of the points, NGC~4565
($\Delta\mu = -0.80$~mag) and NGC~891 ($\Delta\mu =+0.71$~mag) are edge-on
spirals, the only such systems in the sample.  Since SBF measurements to
late-type galaxies are difficult (due to the presence of dust and color
gradients), while photometry of halo PNe is relatively easy, it is tempting
to blame this discrepancy on the SBF values.  However at this time,
all we can say is that one (or both) methods have trouble measuring
distances to these systems.

If the outlying galaxies are excluded, then the scatter in the measurements
exactly follows that predicted from the internal errors of the methods.
This agreement is striking, and strongly suggests that the quoted errors
of the measurements are accurate.  The agreement also leaves little room
for additional sources of error; if there is some population-dependent
term in the scatter, it must be small.

This latter conclusion is confirmed in Figure~8.  If either method were
significantly affected by population age or metallicity, then $\Delta\mu$
would correlate with galactic absolute magnitude or color.  As Figure~8
illustrates, it does not.  Similarly, if the form of the PNLF (equation~2)
were incorrect, $\Delta\mu$ would correlate with absolute magnitude or
PN population.  This correlation does not exist, either.  In fact, the
only possible trend present in the figure occurs when the PNLF-SBF residuals
are plotted against distance.  If one {\it only\/} considers galaxies with
$(m-M)_{\rm SBF} > 30.6$, then there is a 95\% chance that a correlation exists
between $\Delta\mu$ and distance modulus.  Such a trend might be expected
if the PN candidates found in these systems are contaminated by background
emission-line galaxies or (in the case of galaxies in rich clusters)
foreground intracluster stars.  However, if the five most distant objects
are deleted from the sample, the correlation with distance goes away.  Thus,
the overall impression left from the figure is that in terms of relative
distances, the PNLF and SBF techniques are in excellent agreement.

Unfortunately, the same cannot be said for the methods' absolute distances.  As
Figure~7 demonstrates, there is a 0.30~mag offset between the SBF and PNLF
distance scales, in the sense that SBF distances are systematically larger than
their PNLF counterparts.  This offset, which is reduced to $0.26$~mag
if the 5 most distant galaxies are excluded, is highly significant.  Based
on the 13 galaxies with Cepheid and PNLF measurements, the uncertainty in the
PNLF zero point is $\sim 0.05$~mag.  Similarly, the uncertainty in the SBF
zero point, as defined by the six Cepheid calibrators, is $\sim 0.04$~mag.
Finally, there is the uncertainty in the zero point of Galactic extinction.
This continues to be a controversial topic \citep[see][for a review]{bh82},
and, indeed, the zero point of the \citet{schlegel} extinction map is
0.02~mag smaller in $E(B-V)$ than that of the \citet{bh82} map.  However,
the sense of this offset is that, if the \citet{bh82} zero point were adopted,
the discrepancy between the PNLF and SBF distance scales would be worse.  In
fact, if one wishes to avoid negative DIRBE/IRAS reddenings, then the
extinction zero point cannot be lowered by more than $E(B-V)=0.015$~mag.
This translates into a hard upper limit of 0.10~mag on the uncertainty
in the mean value of $\Delta\mu$ due to dust, and a total upper limit on
the uncertainty of 0.12~mag.  The observed offset between the PNLF and
SBF distance scales is therefore statistically significant at more than the
$2 \, \sigma$ level.  Considering the fact that both the SBF and PNLF methods
are calibrated in the same way via Cepheids, this is a remarkable result!

\clearpage

\section{Discussion}
Logically speaking, there are only three possible ways to explain the
discrepancy between the PNLF and SBF distance scales:

{\it The Cepheid distances of the SBF calibrators may be systematically
different than those of the PNLF calibrators.}  This solution is extremely
unlikely, as the two samples of galaxies are nearly identical.  Specifically,
four of the six SBF calibrators are also PNLF calibrators, the median
metallicity of the PNLF calibrators is very nearly the same as that of the
SBF calibrators (O/H $\sim 8.88$ vs.~O/H $\sim 8.80$), and there is no
significant difference in the Cepheid internal reddenings of the two samples.
Consequently, it is difficult to blame the PNLF-SBF discrepancy on the
Cepheids.

{\it The PNLF measurements made in Cepheid galaxies may be systematically
different from those made in SBF galaxies.}  PN identifications are best made
in the spheroidal components of galaxies, where contamination from H~II
regions is minimal.  However, in order to make direct comparisons with the
Pop~I distance indicators, PNLF observations must be performed in galactic
disks.  Thus, one can attempt to explain the PNLF-SBF discrepancy via the
hypothesis that the PNLF cutoff is fainter in spiral disks than in bulges and
halos.  Such an idea is plausible since internal extinction could, in theory,
dim disk planetary nebulae more than bulge objects.

We can test for such an effect by dividing the PNLF calibrators into
two groups: those whose PN surveys included the galaxies' Pop~I components
(the LMC, the SMC, NGC~300, NGC~2403, NGC~3351, NGC~4258, NGC~5253,
and M101), and those whose observations were confined primarily to the bulge
and halo (M31, M81, NGC~3368, and NGC~3627).  When we do this, we find that
there is no significant difference between the two samples of galaxies: the
zero point found in young populations is only $\sim 0.05$~mag fainter than
that observed for old systems.  Moreover, planetary nebula surveys in the
bulge, disk, and halo of M31 \citep{hui94} and M81 \citep{magrini81} have
also failed to find change in the PNLF zero point with stellar population.
It is therefore extremely unlikely that PNLF differences can explain the
0.3~mag offset.

{\it The SBF measurements made in Cepheid galaxies may be systematically
different than those made in PNLF galaxies.}  Of the 28 galaxies with both
PNLF and SBF measurements, all but eight are elliptical or lenticular.  Yet
the SBF calibration is derived from the observation of six spiral bulges.
Thus, the potential exists for a systematic error between the two data
samples.  Indeed, due to the extreme color sensitivity of the SBF method,
a moderate amount of internal extinction ($E(B-V) \sim 0.06$) in the six
calibrating galaxies will shift the SBF scale by more than 10\% and bring
it and bring it into close agreement with the scale defined by the PNLF{}.

Is there evidence for this amount of internal extinction?  It is true that
the direct measurements of extinction in the bulges of M31 \citep{mcclure}
and M81 \citep{peimbert2} are greater than the galaxies' DIRBE/IRAS reddening
values \citep{schlegel}.  However, in both cases the excess is less than
what is needed, $E(B-V) \sim 0.03$.  Similarly, if one breaks the sample of
galaxies in two, one finds a slight difference between the PNLF-SBF distance
offset for spiral bulges ($-0.16 \pm 0.24$~mag) and for ellipticals
($-0.34 \pm 0.19$~mag).  Unfortunately, even if this result is
significant, it still does not fully explain the 0.3~mag offset of
Figure~7.

Nevertheless, internal extinction is still the most likely cause of the
scale discrepancy.  If the bulges of the six Cepheid calibrators have as
little as $E(B-V) \sim 0.04$ of internal extinction, then the zero point of
the SBF system shifts by 0.14~mag and the SBF-based Hubble Constant
of \citet{ferrarese} moves from $H_0=69$ to $H_0 = 74$~km~s$^{-1}$~Mpc$^{-1}$.
[Note that this shift is almost as large as the formal
$\pm 6$~km~s$^{-1}$~Mpc$^{-1}$ systematic error assigned by \citet{ferrarese}
and \citet{keyfinal}.]  Moreover, if the same extinction is applied to the
PNLF measurements, then the PNLF scale is increased by a comparable amount, and
the PNLF/SBF discrepancy disappears completely.  This result underscores
the need close study of the internal extinction of galaxies, and for more
independent crosschecks at each rung of the distance ladder.

\section{The PNLF-TRGB Comparison}
Note that our conclusion differs from that of \citet{ferrarese}, who
attributed the PNLF/SBF scale discrepancy to a systematic error in the
planetary nebula distances.  The evidence \citet{ferrarese} cited in
in support of this interpretation was 1) the similarity between the SBF
distances to the ellipticals of Virgo and Fornax, and the Cepheid distances
to the clusters' spirals, and 2) the agreement between the SBF distance
scale, and the scale inferred from $I$-band measurements of stars at the
the tip of the red giant branch (TRGB\null).  The former argument is
suspect, since both clusters are complex, and the elliptical galaxies are
not co-mingled with the spirals \citep{wb00, dgc01}.  The second comment,
however, deserves some attention.

The TRGB method is well tested inside the Local Group: the data from nine
nearby galaxies with both TRGB and Cepheid measurements demonstrate that the
technique can produce results that are accurate to better than $\sim 0.2$~mag
\citep{ferrarese}.  However, no Cepheid galaxy more than 3~Mpc away has a TRGB
measurement: at these distances, the only direct comparisons are with PNLF and
SBF galaxies.

A galaxy-by-galaxy comparison of TRGB, SBF, and PNLF distances appears in
Figure~9.  As the figure illustrates, the overlap between TRGB, PNLF, and
SBF galaxies is scant.  Only 7 PNLF galaxies have TRGB distances: the
LMC \citep{szk}, M31 \citep{durrell}, M33 \citep{kim}, NGC~5102 \citep{kara},
NGC~5128 \citep{harris}, NGC~3379 \citep{sakai}, and NGC~3115 \citep{elson}.
Only five galaxies have both TRGB and SBF measurements (M31, NGC~5102,
NGC~5128, NGC~3379, and NGC~3115).  If one just considers the mean offsets
between the measurements, then the data do indicate that the SBF and TRGB
distance scales are in agreement, and that the PNLF scale is $\sim -0.2$~mag
too small.  However, Figure~9 contains two other properties that should not
be ignored.

The first concerns the size of the SBF-TRGB internal error bars relative to
the observed scatter in the measurements.  An important conclusion of \S 5
and 6 is that the scatter between the PNLF, SBF, and Cepheid distance
measurements is consistent with that expected from the internal errors of
the methods.  This agreement strongly suggests that the quoted errors for
all three methods are accurate.  However, the same is not true for the
SBF-TRGB comparison.  As Figure~9 demonstrates, the observed SBF-TRGB distance
residuals are significantly larger than the internal errors of the methods
would predict. (The internal and external errors are incompatible at the
$\sim 80\%$ confidence level.)  Since the SBF errors appear to be reasonable,
the implication is that the TRGB measurements possess an additional source of
uncertainty.

The second property displayed in Figure~9 is the trend exhibited in the
PNLF-TRGB comparison.  Inside the Local Group, there is excellent agreement
between the PNLF and TRGB distance indicators; the mean difference
between the two methods is $-0.01 \pm 0.06$~mag.  At larger distances,
however, the techniques diverge, and $\Delta\mu = -0.29 \pm 0.05$.
Since the Cepheid and SBF comparisons demonstrate there is no problem
with the relative PNLF distances (at least for objects within $\sim 10$~Mpc),
the existence of the discrepancy again suggests a problem with the TRGB
data.

With the information currently available, we cannot prove that an error
exists in the TRGB distance measurements.  However, we can speculate on
the types of errors which may be associated with the method.  TRGB
measurements require that the red giant branch be well-populated:  if the
RGB is underpopulated, then the apparent magnitude of the RGB tip will be
overestimated, and the distance to the parent galaxy will be overestimated
\citep{mf95}.  Even more important is the effect of metallicity.  Even if
the sample of RGB stars is large, the technique still will not work unless
the stars at the RGB tip have [Fe/H] $\lesssim -0.7$ \citep{lee93, bellazzini}.
If the red giants are more metal-rich than this, their absolute $I$-band
luminosities will be fainter than assumed, due to the effects of line
blanketing, and the result will again be an overestimate of distance.
Thus, the systematic errors associated with the technique are asymmetric:
if an error exists, then the distances produced by the TRGB technique will
be upper limits.\footnote{\citet{mf95} note that image crowding and low
signal-to-noise detections can potentially bias TRGB measurements towards
brighter magnitudes (smaller distances).  However, these effects will
almost always be small ($\lesssim 0.1$~mag).  The systematic effects of
sample size and metallicity are much larger.}

For nearby galaxies, the metallicity effect on the TRGB is not of great
concern.  Most Local Group galaxies are small and thus metal-poor; the
only exceptions (M31 and M33) both have high quality data and measured
RGB metallicity distributions.  Consequently, the TRGB distances to these
objects are well-determined, and in excellent agreement with the distances
derived from Cepheid \citep{ferrarese}, PNLF, and SBF measurements.
Outside the Local Group, however, the data quality is generally poorer, and
with one exception (NGC~5128), little or no information is available about
the RGB metallicity distribution.  Since most of these distant galaxies are
more massive (and therefore more metal-rich) than their Local Group
counterparts, a systematic error in the TRGB distance scale is not
inconceivable.   Thus, the PNLF and TRGB measurement may, indeed, be
consistent.  Higher quality, multicolor TRGB measurements beyond the
Local Group are needed to resolve the problem.

\section{Conclusion}
We have presented PNLF distances to NGC~2403, NGC~3115, NGC~3351, NGC~3627,
NGC~4258, and NGC~5866, and have used these data to compare the Cepheid,
PNLF, and SBF distance scales.  Our observations demonstrate that, in
terms of relative distance measurements, the Cepheid, PNLF, and SBF
methods are in excellent agreement, and the internal errors estimated
for all the methods are correct.  However, we also show that the PNLF and
SBF distance scales are incompatible:  the Cepheid-calibrated SBF scale
is $\sim 0.3$~mag longer than the Cepheid-calibrated PNLF scale.  The likely
cause of the discrepancy is internal extinction in the bulges of the SBF
Cepheid calibrators.  If this is true, then this error results in an
underestimate of the SBF Hubble Constant.  Finally, we use our PNLF distance
to NGC~4258, in combination with the galaxy's geometric and Cepheid distances,
to argue that the short distance to the Large Magellanic Cloud is correct, and
that the \citet{keyfinal} Hubble Constant should be increased by $8 \pm 3\%$.

\acknowledgements
We would like to that R. M\'endez for the list of PN magnitudes in NGC~4697.
We would also like to thank the anonymous referee for suggesting that we
revisit the TRGB distance calibration.  This research made use of the NASA
Extragalactic Database and was supported in part by NSF grants AST 95-29270
and AST 00-71238.

\clearpage

\begin{deluxetable}{lcccccccccc}
\rotate
\tabletypesize\footnotesize
\setlength{\tabcolsep}{0.075in}
\tablewidth{0pt}
\tablecaption{Observing Log}
\tablehead{
&&&&\colhead{Image Scale}
&&&\multicolumn{2}{c}{Exp time (min)}
&&\colhead{Limiting} \\
\colhead{Galaxy}
&\colhead{Telescope}
&\colhead{Detector}
&\colhead{Field-of-View}
&\colhead{(per pixel)}
&\colhead{Filter}
&\colhead{Date}
&\colhead{[O~III] $\lambda 5007$} \
&\colhead{H$\alpha$}
&\colhead{Seeing}
&\colhead{$m_{5007}$} \\
}
\startdata
M31's bulge &KPNO 2.1~m &T1KA &$5\farcm 3 \times 5\farcm 3$ &$0\farcs 31$
&KP1413 &1994 Jan &75 &45 &$0\farcs 9$ &24.0 \\
NGC 2403 &KPNO 4~m &T2KB &$16\farcm 4 \times 16\farcm 4$ &$0\farcs 48$
&KP1389  &1996 Nov &135 &45 &$1\farcs 3$ &24.1 \\
NGC 3115 North &KPNO 4~m &RCA3 &$5\farcm 1 \times 3\farcm 0$ &$0\farcs 60$
&5016/28 &1985 Mar &180 &\nodata &$1\farcs 1$ &26.2 \\
NGC 3115 South &KPNO 4~m &RCA3 &$5\farcm 1 \times 3\farcm 0$ &$0\farcs 60$
&5016/28 &1985 Mar &240 &\nodata &$1\farcs 3$ &26.2 \\
NGC 3351 &WIYN &MiniMo &$4\farcm 8 \times 4\farcm 8$ &$0\farcs 14$
&5027/30 &2001 Mar &180 &45 &$0\farcs 9$ &26.3 \\
NGC 3627 &KPNO 4~m &T2KB  &$16\farcm 4 \times 16\farcm 4$ &$0\farcs 48$
&5027/30 &1997 Mar &240 &45 &$1\farcs 3$ &26.2 \\
NGC 4258 &WIYN &MiniMo &$9\farcm 6 \times 9\farcm 6$ &$0\farcs 14$
&KP1590  &2001 Mar &120 &40 &$1\farcs 0$ &25.6 \\
NGC 5866 &KPNO 4~m &TI2 &$4\farcm 0 \times 4\farcm 0$ &$0\farcs 60$
&5016/28 &1985 Apr &180 &\nodata &$1\farcs 2$ &26.7 \\
\enddata
\end{deluxetable}

\begin{deluxetable}{lcccr}
\tablewidth{0pt}
\tablecaption{M31 Planetaries}
\tablehead{
\colhead{ID} &\colhead{$\alpha(2000)$} &\colhead{$\delta(2000)$}
&\colhead{$m_{5007}$} &\colhead{R} \\
}
\startdata
  1   &0 42 46.10   &41 16 40.9   &20.51   &3.5   \\
  2   &0 42 45.93   &41 16 24.3   &21.84   &0.9   \\
  3   &0 42 43.73   &41 16 25.7   &20.85   &3.3   \\
  5   &0 42 42.34   &41 15 53.7   &21.07   &3.3   \\
  7   &0 42 43.27   &41 15 56.5   &21.62   &1.5   \\
  8   &0 42 43.86   &41 16 01.1   &21.49   &2.1   \\
  9   &0 42 45.15   &41 16 05.4   &21.28   &2.4   \\
 10   &0 42 46.66   &41 16 09.7   &21.12   &2.2   \\
 12   &0 42 47.50   &41 16 21.3   &20.70   &3.0   \\
 13   &0 42 47.25   &41 16 28.7   &21.48   &2.1   \\
 14   &0 42 46.88   &41 16 55.7   &22.28   &1.4   \\
 15   &0 42 46.35   &41 17 02.0   &21.43   &1.0   \\
 16   &0 42 40.91   &41 16 11.7   &21.83   &3.8   \\
 17   &0 42 39.74   &41 15 49.2   &20.71   &2.8   \\
 18   &0 42 39.72   &41 15 36.6   &20.88   &2.2   \\
 20   &0 42 43.05   &41 15 36.8   &21.63   &2.0   \\
 21   &0 42 45.15   &41 15 23.6   &21.11   &3.5   \\
 23   &0 42 48.89   &41 16 55.5   &21.34   &2.9   \\
 24   &0 42 53.31   &41 16 27.9   &21.28   &3.0   \\
 25   &0 42 54.60   &41 16 24.3   &21.73   &2.8   \\
 26   &0 42 55.44   &41 16 24.4   &21.17   &3.0   \\
 28   &0 42 52.06   &41 17 24.5   &20.65   &3.2   \\
 29   &0 42 53.48   &41 17 33.9   &21.01   &3.8   \\
 30   &0 42 55.40   &41 17 20.8   &20.70   &2.1   \\
 32   &0 42 53.32   &41 18 17.1   &20.79   &3.3   \\
 35   &0 42 38.27   &41 15 33.8   &21.35   &3.4   \\
 36   &0 42 37.38   &41 15 51.2   &20.94   &2.3   \\
 37   &0 42 37.21   &41 16 26.7   &22.05   &3.0   \\
 38   &0 42 37.72   &41 16 43.3   &21.37   &1.8   \\
 39   &0 42 36.40   &41 16 57.4   &21.76   &3.2   \\
 40   &0 42 36.02   &41 16 34.4   &21.51   &2.8   \\
 41   &0 42 33.21   &41 16 49.4   &20.65   &3.6   \\
 42   &0 42 32.74   &41 16 32.7   &20.43   &3.7   \\
 43   &0 42 31.00   &41 16 25.3   &21.20   &2.6   \\
 44   &0 42 31.45   &41 16 14.7   &21.99   &2.6   \\
 45   &0 42 32.50   &41 15 58.5   &20.59   &2.9   \\
 51   &0 42 34.13   &41 15 04.8   &21.02   &3.6   \\
 52   &0 42 34.69   &41 14 44.8   &21.64   &2.5   \\
 53   &0 42 35.25   &41 14 46.2   &20.40   &3.3   \\
 54   &0 42 37.08   &41 14 35.5   &20.61   &4.0   \\
 55   &0 42 38.38   &41 14 34.6   &21.34   &3.3   \\
 56   &0 42 40.69   &41 14 10.0   &20.97   &2.8   \\
 57   &0 42 42.10   &41 14 09.5   &21.10   &3.3   \\
 58   &0 42 43.84   &41 14 50.7   &21.13   &1.7   \\
 59   &0 42 46.04   &41 15 16.6   &21.60   &2.2   \\
 60   &0 42 47.50   &41 15 05.2   &21.49   &1.8   \\
 61   &0 42 46.71   &41 14 21.4   &20.93   &3.3   \\
 62   &0 42 48.89   &41 15 24.0   &21.09   &3.2   \\
 63   &0 42 49.33   &41 14 56.7   &21.76   &3.8   \\
 64   &0 42 52.69   &41 14 15.7   &20.78   &3.7   \\
 65   &0 42 39.39   &41 14 17.6   &22.07   &4.2   \\
 66   &0 42 40.07   &41 14 38.3   &21.97   &1.2   \\
 70   &0 42 38.96   &41 14 56.5   &21.83   &1.9   \\
 71   &0 42 37.49   &41 14 35.0   &22.00   &1.4   \\
 72   &0 42 46.34   &41 15 46.3   &21.40   &4.2   \\
 74   &0 42 44.88   &41 15 21.0   &21.71   &1.9   \\
 76   &0 42 54.56   &41 15 37.6   &21.91   &2.5   \\
 77   &0 42 54.56   &41 15 21.9   &21.96   &3.1   \\
 78   &0 42 50.89   &41 14 47.6   &21.93   &0.9   \\
 79   &0 42 48.62   &41 14 26.3   &22.27   &1.1   \\
 80   &0 42 57.30   &41 17 25.5   &20.98   &3.2   \\
 81   &0 42 43.38   &41 16 57.5   &22.53   &1.8   \\
 82   &0 42 39.76   &41 17 03.3   &22.48   &1.3   \\
 83   &0 42 41.35   &41 17 49.5   &22.30   &3.7   \\
 87   &0 42 38.04   &41 16 52.8   &22.11   &1.3   \\
 99   &0 42 42.43   &41 13 56.5   &22.10   &1.2   \\
102   &0 42 39.51   &41 18 31.4   &22.62   &1.6   \\
103   &0 42 40.16   &41 18 41.9   &22.63   &1.3   \\
112   &0 42 49.63   &41 18 41.0   &22.76   &2.2   \\
120   &0 42 56.07   &41 17 32.6   &22.22   &1.2   \\
136   &0 42 57.16   &41 16 59.2   &22.11   &2.0   \\
137   &0 42 50.10   &41 15 26.7   &22.26   &1.1   \\
139   &0 42 43.07   &41 14 09.3   &22.53   &0.9   \\
141   &0 42 41.34   &41 14 07.4   &22.08   &2.8   \\
142   &0 42 32.86   &41 15 58.2   &21.88   &2.6   \\
143   &0 42 38.36   &41 14 27.7   &22.53   &2.8   \\
160   &0 42 40.21   &41 13 51.0   &22.82   &1.8   \\
164   &0 42 49.26   &41 13 53.6   &22.24   &3.0   \\
165   &0 42 47.56   &41 14 59.8   &22.66   &2.4   \\
166   &0 42 49.00   &41 14 42.7   &22.61   &2.1   \\
171   &0 42 56.89   &41 14 08.8   &23.32   &2.5   \\
174   &0 42 42.14   &41 14 22.7   &22.57   &3.4   \\
175   &0 42 55.76   &41 16 16.1   &22.12   &2.5   \\
176   &0 42 54.19   &41 15 28.0   &23.05   &1.3   \\
181   &0 42 40.33   &41 14 10.0   &21.97   &2.8   \\
189   &0 42 57.36   &41 17 14.8   &22.87   &1.0   \\
190   &0 42 55.31   &41 18 15.0   &22.67   &1.4   \\
191   &0 42 56.91   &41 18 14.4   &23.11   &2.9   \\
316   &0 42 44.23   &41 16 03.7   &21.04   &2.9   \\
323   &0 42 45.24   &41 15 29.6   &22.15   &1.2   \\
328   &0 42 49.34   &41 16 12.3   &22.29   &0.7   \\
330   &0 42 49.58   &41 16 50.6   &22.29   &1.0   \\
420   &0 42 30.29   &41 14 07.7   &23.35   &0.5   \\
421   &0 42 30.71   &41 17 53.3   &23.38   &3.4   \\
423   &0 42 31.33   &41 14 25.9   &23.37   &2.1   \\
424   &0 42 31.95   &41 17 32.4   &23.14   &2.5   \\
427   &0 42 34.09   &41 16 48.4   &23.73   &0.6   \\
428   &0 42 34.10   &41 16 31.2   &22.67   &1.8   \\
431   &0 42 35.19   &41 17 27.9   &23.04   &0.6   \\
433   &0 42 35.78   &41 13 50.3   &22.72   &1.6   \\
434   &0 42 35.68   &41 18 23.3   &23.44   &0.4   \\
440   &0 42 38.67   &41 14 09.9   &23.09   &1.2   \\
441   &0 42 38.96   &41 13 59.7   &23.71   &0.1   \\
442   &0 42 39.99   &41 16 20.0   &23.43   &0.6   \\
443   &0 42 40.03   &41 15 52.3   &23.03   &3.0   \\
445   &0 42 40.56   &41 17 24.7   &23.03   &1.9   \\
447   &0 42 41.26   &41 18 21.0   &23.21   &1.9   \\
449   &0 42 41.51   &41 15 01.5   &22.65   &2.3   \\
453   &0 42 44.45   &41 18 13.0   &23.34   &0.5   \\
454   &0 42 45.79   &41 16 01.5   &22.62   &1.1   \\
455   &0 42 45.97   &41 18 24.8   &22.92   &1.6   \\
460   &0 42 48.12   &41 17 33.9   &22.94   &2.9   \\
462   &0 42 51.31   &41 15 56.6   &21.94   &3.2   \\
464   &0 42 50.53   &41 18 35.6   &23.60   &1.5   \\
465   &0 42 50.74   &41 17 36.1   &24.31   &3.3   \\
466   &0 42 51.71   &41 16 04.7   &23.41   &0.8   \\
467   &0 42 51.94   &41 17 05.4   &23.81   &4.0   \\
468   &0 42 52.22   &41 15 53.3   &23.22   &1.2   \\
471   &0 42 53.00   &41 17 15.9   &24.50   &0.5   \\
472   &0 42 53.69   &41 13 59.8   &23.25   &1.5   \\
473   &0 42 53.90   &41 17 02.4   &23.06   &0.4   \\
474   &0 42 54.07   &41 15 04.3   &23.29   &0.7   \\
476   &0 42 54.21   &41 14 17.6   &22.83   &3.2   \\
477   &0 42 55.03   &41 17 02.4   &22.86   &1.7   \\
478   &0 42 55.23   &41 17 12.1   &22.09   &0.8   \\
570   &0 42 55.60   &41 16 23.8   &22.12   &2.4   \\
571   &0 42 43.02   &41 15 32.2   &22.29   &1.1   \\
572   &0 42 42.92   &41 16 16.3   &22.33   &1.6   \\
573   &0 42 44.79   &41 16 59.3   &22.39   &1.5   \\
574   &0 42 44.97   &41 16 04.1   &22.44   &1.5   \\
575   &0 42 46.27   &41 15 23.0   &22.46   &1.5   \\
576   &0 42 42.90   &41 15 20.2   &22.52   &1.8   \\
577   &0 42 43.10   &41 16 40.7   &22.72   &1.5   \\
578   &0 42 48.69   &41 16 09.6   &22.72   &2.5   \\
579   &0 42 37.99   &41 16 02.4   &22.75   &3.3   \\
580   &0 42 54.22   &41 17 58.1   &22.75   &1.9   \\
581   &0 42 44.38   &41 15 05.3   &22.75   &3.2   \\
582   &0 42 47.56   &41 18 51.3   &22.85   &1.8   \\
583   &0 42 47.69   &41 17 24.3   &22.89   &2.5   \\
584   &0 42 41.70   &41 14 28.4   &22.91   &1.8   \\
585   &0 42 43.84   &41 17 00.1   &22.97   &0.8   \\
586   &0 42 42.12   &41 15 35.6   &22.99   &0.7   \\
587   &0 42 51.06   &41 17 06.0   &23.01   &2.3   \\
588   &0 42 35.38   &41 14 44.4   &23.03   &1.8   \\
589   &0 42 44.55   &41 18 50.2   &23.04   &0.4   \\
590   &0 42 43.30   &41 17 13.1   &23.06   &0.6   \\
591   &0 42 54.35   &41 14 59.6   &23.06   &3.0   \\
592   &0 42 45.98   &41 14 26.4   &23.06   &1.1   \\
593   &0 42 33.98   &41 15 02.8   &23.06   &1.0   \\
594   &0 42 43.29   &41 16 52.5   &23.07   &4.7   \\
595   &0 42 47.24   &41 15 51.2   &23.07   &0.8   \\
596   &0 42 40.15   &41 15 22.3   &23.08   &1.5   \\
597   &0 42 41.55   &41 15 58.2   &23.09   &1.1   \\
598   &0 42 45.49   &41 15 31.8   &23.10   &1.2   \\
599   &0 42 47.02   &41 17 07.0   &23.10   &2.2   \\
600   &0 42 49.30   &41 17 29.7   &23.15   &$> 4.5$ \\
601   &0 42 43.21   &41 14 41.2   &23.16   &1.3   \\
602   &0 42 49.77   &41 16 25.0   &23.16   &1.2   \\
603   &0 42 54.58   &41 17 16.1   &23.17   &2.2   \\
604   &0 42 31.84   &41 16 03.3   &23.18   &2.6   \\
605   &0 42 35.95   &41 14 10.7   &23.19   &0.5   \\
606   &0 42 49.44   &41 16 31.9   &23.22   &1.4   \\
607   &0 42 44.08   &41 15 51.3   &23.23   &0.4   \\
608   &0 42 48.24   &41 18 57.8   &23.23   &1.8   \\
609   &0 42 39.09   &41 16 34.4   &23.23   &1.8   \\
610   &0 42 42.32   &41 16 24.7   &23.23   &1.0   \\
611   &0 42 43.80   &41 15 39.3   &23.26   &0.7   \\
612   &0 42 46.69   &41 14 57.2   &23.26   &2.3   \\
613   &0 42 38.97   &41 15 48.1   &23.27   &1.8   \\
614   &0 42 47.97   &41 14 59.0   &23.29   &0.5   \\
615   &0 42 42.06   &41 15 37.5   &23.29   &0.8   \\
616   &0 42 46.67   &41 18 20.4   &23.31   &0.7   \\
617   &0 42 41.50   &41 16 20.0   &23.32   &0.8   \\
618   &0 42 45.35   &41 18 44.0   &23.32   &2.2   \\
619   &0 42 54.68   &41 18 56.0   &23.33   &1.8   \\
620   &0 42 40.49   &41 15 01.5   &23.33   &0.7   \\
621   &0 42 40.27   &41 15 44.3   &23.34   &1.9   \\
622   &0 42 42.63   &41 16 06.0   &23.34   &0.3   \\
623   &0 42 40.68   &41 16 15.9   &23.35   &3.8   \\
624   &0 42 35.19   &41 15 03.2   &23.35   &0.5   \\
625   &0 42 55.03   &41 17 41.4   &23.37   &0.4   \\
626   &0 42 39.69   &41 15 57.6   &23.37   &2.7   \\
627   &0 42 47.36   &41 15 46.1   &23.38   &0.5   \\
628   &0 42 52.01   &41 18 07.9   &23.39   &4.0   \\
629   &0 42 44.51   &41 18 52.0   &23.39   &$>3.6$ \\
630   &0 42 45.88   &41 18 46.0   &23.40   &1.8   \\
631   &0 42 40.07   &41 15 07.9   &23.41   &1.3   \\
632   &0 42 38.45   &41 18 52.5   &23.42   &0.7   \\
633   &0 42 45.78   &41 17 35.2   &23.42   &0.4   \\
634   &0 42 51.44   &41 15 51.3   &23.45   &1.7   \\
635   &0 42 43.93   &41 18 38.5   &23.47   &3.1   \\
636   &0 42 37.50   &41 18 00.4   &23.51   &1.5   \\
637   &0 42 51.49   &41 16 19.8   &23.51   &2.4   \\
638   &0 42 35.98   &41 15 27.0   &23.51   &1.1   \\
639   &0 42 40.85   &41 15 18.5   &23.53   &1.1   \\
640   &0 42 49.15   &41 17 25.6   &23.54   &$>3.2$  \\
641   &0 42 33.00   &41 14 34.4   &23.55   &1.5   \\
642   &0 42 42.12   &41 18 46.0   &23.55   &1.7   \\
643   &0 42 42.07   &41 14 40.6   &23.56   &1.6   \\
644   &0 42 40.16   &41 15 32.9   &23.57   &0.8   \\
645   &0 42 45.65   &41 14 21.0   &23.58   &2.2   \\
646   &0 42 35.42   &41 15 45.5   &23.60   &1.0   \\
647   &0 42 32.24   &41 16 44.3   &23.60   &1.0   \\
648   &0 42 51.43   &41 18 39.9   &23.60   &0.2   \\
649   &0 42 51.76   &41 17 45.1   &23.65   &0.2   \\
650   &0 42 31.73   &41 14 41.5   &23.65   &2.0   \\
651   &0 42 45.07   &41 17 10.4   &23.68   &0.6   \\
652   &0 42 40.06   &41 14 05.5   &23.68   &0.8   \\
653   &0 42 51.83   &41 14 25.0   &23.71   &3.4   \\
654   &0 42 46.56   &41 15 32.8   &23.74   &0.1   \\
655   &0 42 43.67   &41 18 07.7   &23.75   &1.4   \\
656   &0 42 53.32   &41 18 27.3   &23.75   &0.7   \\
657   &0 42 51.87   &41 15 16.8   &23.75   &1.8   \\
658   &0 42 47.82   &41 15 41.9   &23.77   &1.4   \\
659   &0 42 43.89   &41 17 32.7   &23.77   &1.8   \\
660   &0 42 51.72   &41 18 15.2   &23.78   &5.4   \\
661   &0 42 35.22   &41 15 52.9   &23.79   &0.5   \\
662   &0 42 48.44   &41 13 57.2   &23.79   &1.6   \\
663   &0 42 47.93   &41 14 42.7   &23.79   &1.5   \\
664   &0 42 43.46   &41 15 08.6   &23.80   &0.6   \\
665   &0 42 31.10   &41 13 50.7   &23.80   &1.2   \\
666   &0 42 46.13   &41 18 40.0   &23.80   &1.7   \\
667   &0 42 41.77   &41 16 38.2   &23.81   &1.8   \\
668   &0 42 55.11   &41 14 29.8   &23.82   &0.3   \\
669   &0 42 47.04   &41 15 22.5   &23.83   &1.3   \\
670   &0 42 47.79   &41 18 50.8   &23.84   &1.8   \\
671   &0 42 55.00   &41 14 27.1   &23.85   &1.0   \\
672   &0 42 37.29   &41 16 55.9   &23.90   &1.2   \\
673   &0 42 45.51   &41 17 37.0   &23.92   &1.9   \\
674   &0 42 39.01   &41 16 36.5   &23.93   &0.7   \\
675   &0 42 37.85   &41 15 24.6   &23.93   &0.5   \\
676   &0 42 51.77   &41 16 31.4   &23.94   &3.1   \\
677   &0 42 47.44   &41 18 38.0   &23.94   &1.0   \\
678   &0 42 44.47   &41 17 14.1   &23.95   &$>2.2$  \\
679   &0 42 53.00   &41 16 37.2   &23.96   &2.7   \\
680   &0 42 52.10   &41 16 35.6   &23.96   &2.1   \\
681   &0 42 36.74   &41 16 18.9   &23.97   &1.5   \\
682   &0 42 55.61   &41 16 18.9   &23.97   &$>2.1$ \\
683   &0 42 51.74   &41 15 51.6   &23.97   &1.9   \\
684   &0 42 36.17   &41 15 15.4   &23.99   &$>2.1$ \\
685   &0 42 50.50   &41 17 52.9   &24.01   &0.9   \\
686   &0 42 47.39   &41 14 27.0   &24.02   &1.1   \\
687   &0 42 52.45   &41 16 14.2   &24.02   &0.4   \\
688   &0 42 48.71   &41 14 14.9   &24.02   &0.9   \\
689   &0 42 34.98   &41 13 54.2   &24.03   &1.5   \\
690   &0 42 44.33   &41 14 22.6   &24.04   &1.2   \\
691   &0 42 33.61   &41 14 35.7   &24.05   &0.5   \\
692   &0 42 34.89   &41 14 32.0   &24.05   &1.5   \\
693   &0 42 39.31   &41 14 29.3   &24.05   &0.2   \\
694   &0 42 39.46   &41 16 07.2   &24.06   &2.0   \\
695   &0 42 43.24   &41 17 53.8   &24.07   &1.2   \\
696   &0 42 51.94   &41 18 22.9   &24.07   &1.7   \\
697   &0 42 56.33   &41 14 27.3   &24.07   &1.1   \\
698   &0 42 50.79   &41 18 17.1   &24.08   &1.1   \\
699   &0 42 39.11   &41 18 33.4   &24.09   &2.1   \\
700   &0 42 56.92   &41 16 42.8   &24.09   &1.8   \\
701   &0 42 52.74   &41 17 50.0   &24.11   &1.4   \\
702   &0 42 32.61   &41 15 15.6   &24.11   &3.0   \\
703   &0 42 31.29   &41 15 01.0   &24.11   &0.5   \\
704   &0 42 33.78   &41 15 37.4   &24.11   &$>1.9$ \\
705   &0 42 43.85   &41 17 47.1   &24.12   &$>1.8$ \\
706   &0 42 47.33   &41 18 03.3   &24.12   &1.0   \\
707   &0 42 39.76   &41 16 41.3   &24.14   &0.1   \\
708   &0 42 31.48   &41 15 58.8   &24.15   &0.6   \\
709   &0 42 43.03   &41 18 09.7   &24.16   &1.1   \\
710   &0 42 41.40   &41 14 53.7   &24.16   &0.2   \\
711   &0 42 50.45   &41 18 08.2   &24.17   &1.3   \\
712   &0 42 36.88   &41 14 19.5   &24.18   &0.7   \\
713   &0 42 39.56   &41 18 54.9   &24.19   &0.6   \\
714   &0 42 38.13   &41 15 09.3   &24.20   &0.6   \\
715   &0 42 52.58   &41 17 07.8   &24.20   &0.6   \\
716   &0 42 33.31   &41 16 07.1   &24.20   &0.8   \\
717   &0 42 33.49   &41 18 02.8   &24.21   &1.5   \\
718   &0 42 46.46   &41 14 24.7   &24.22   &0.7   \\
719   &0 42 37.27   &41 15 41.0   &24.22   &2.2   \\
720   &0 42 55.89   &41 15 18.3   &24.23   &0.6   \\
721   &0 42 47.50   &41 14 03.4   &24.27   &0.5   \\
722   &0 42 36.13   &41 17 20.0   &24.29   &0.5   \\
723   &0 42 35.86   &41 17 52.1   &24.30   &1.7   \\
724   &0 42 37.21   &41 16 47.9   &24.30   &0.8   \\
725   &0 42 34.61   &41 18 31.5   &24.33   &0.1   \\
726   &0 42 46.62   &41 17 23.9   &24.34   &0.7   \\
727   &0 42 54.05   &41 16 59.7   &24.41   &2.4   \\
728   &0 42 30.01   &41 16 35.5   &24.45   &2.0   \\
729   &0 42 38.48   &41 18 07.8   &24.45   &1.3   \\
730   &0 42 36.76   &41 17 42.4   &24.48   &1.2   \\
731   &0 42 53.53   &41 17 16.0   &24.53   &0.7   \\
732   &0 42 48.46   &41 18 21.8   &24.53   &0.7   \\
733   &0 42 31.72   &41 15 51.1   &24.55   &0.7   \\
734   &0 42 35.18   &41 14 22.5   &24.55   &0.6   \\
735   &0 42 35.06   &41 17 07.5   &24.73   &2.0   \\
736   &0 42 55.97   &41 14 41.1   &24.76   &1.9   \\
737   &0 42 53.04   &41 17 21.3   &24.79   &0.7   \\
738   &0 42 32.66   &41 18 48.7   &24.82   &0.8   \\
739   &0 42 34.34   &41 18 40.3   &24.83   &1.9   \\
740   &0 42 30.25   &41 18 55.2   &24.88   &1.0   \\
741   &0 42 34.27   &41 18 23.7   &24.88   &0.5   \\
742   &0 42 56.70   &41 17 55.9   &24.93   &2.3   \\

\enddata
\end{deluxetable}

\begin{deluxetable}{lcccc}
\tablewidth{0pt}
\tablecaption{NGC 2403 Planetaries}
\tablehead{
\colhead{ID} &\colhead{$\alpha(2000)$} &\colhead{$\delta(2000)$}
&\colhead{$m_{5007}$} &\colhead{Sample} \\
}
\startdata
1   &7 36 20.63  &65 39 04.2  &23.30  &S  \\
2   &7 36 37.38  &65 36 52.0  &23.39  &S  \\
3   &7 37 12.54  &65 33 20.6  &23.41  &S  \\
4   &7 36 51.71  &65 35 01.1  &23.56  &   \\
5   &7 36 57.62  &65 34 58.2  &23.60  &S  \\
6   &7 35 48.67  &65 35 06.0  &23.68  &S  \\
7   &7 36 37.23  &65 34 21.6  &23.71  &S  \\
8   &7 36 36.66  &65 31 28.1  &23.73  &S  \\
9   &7 36 58.06  &65 36 16.7  &23.77  &S  \\
10  &7 37 05.22  &65 34 04.4  &23.78  &S  \\
11  &7 36 23.15  &65 34 58.6  &23.79  &S  \\
12  &7 36 57.97  &65 34 28.7  &23.84  &S  \\
13  &7 36 03.43  &65 40 31.2  &23.85  &S  \\
14  &7 36 05.00  &65 36 13.6  &23.88  &S  \\
15  &7 36 30.93  &65 37 36.0  &23.89  &S  \\
16  &7 37 01.62  &65 35 32.4  &23.90  &S  \\
17  &7 36 03.72  &65 38 29.5  &23.91  &S  \\
18  &7 36 39.47  &65 35 00.2  &23.93  &S  \\
19  &7 36 39.47  &65 36 51.7  &23.98  &S  \\
20  &7 36 19.46  &65 35 33.3  &23.98  &S  \\
21  &7 36 08.88  &65 36 59.4  &24.02  &S  \\
22  &7 36 52.60  &65 34 31.2  &24.11  &   \\
23  &7 36 19.32  &65 33 05.4  &24.15  &   \\
24  &7 37 29.69  &65 32 13.2  &24.19  &   \\
25  &7 36 49.78  &65 37 11.0  &24.21  &   \\
26  &7 37 45.77  &65 31 52.6  &24.23  &   \\
27  &7 37 20.06  &65 40 26.3  &24.42  &   \\
28  &7 35 37.31  &65 38 49.0  &24.46  &   \\
29  &7 37 29.64  &65 31 19.2  &24.48  &   \\
30  &7 37 12.34  &65 33 10.4  &24.50  &   \\
31  &7 36 48.42  &65 36 58.0  &24.51  &   \\
32  &7 35 51.93  &65 40 00.8  &24.74  &   \\
33  &7 36 17.08  &65 37 00.8  &24.80  &   \\
34  &7 36 03.76  &65 42 28.9  &24.91  &   \\
35  &7 36 12.87  &65 39 21.5  &24.93  &   \\
36  &7 37 23.18  &65 37 28.3  &25.21  &   \\
37  &7 37 43.87  &65 35 08.2  &25.47  &   \\
38  &7 36 41.22  &65 40 41.0  &25.68  &   \\
39  &7 36 59.74  &65 29 37.7  &25.68  &   \\
40  &7 37 18.75  &65 30 01.9  &25.72  &   \\
\enddata
\end{deluxetable}

\clearpage

\begin{deluxetable}{lcccc}
\tablewidth{0pt}
\tablecaption{NGC 3115 Planetaries}
\tablehead{
\colhead{ID} &\colhead{$\alpha(2000)$} &\colhead{$\delta(2000)$}
&\colhead{$m_{5007}$} &\colhead{Sample} \\
}
\startdata

 1  &10 05 09.71 &-7 44 05.0  &25.29  &   \\
 2  &10 05 14.54 &-7 41 52.8  &25.40  &S  \\
 3  &10 05 13.61 &-7 43 53.7  &25.40  &   \\
 4  &10 05 08.09 &-7 44 03.7  &25.58  &S  \\
 5  &10 05 14.49 &-7 42 34.4  &25.68  &   \\
 6  &10 05 18.25 &-7 40 04.5  &25.69  &S  \\
 7  &10 05 14.50 &-7 42 18.2  &25.73  &   \\
 8  &10 05 10.99 &-7 43 27.2  &25.73  &   \\
 9  &10 05 09.65 &-7 44 39.7  &25.79  &S  \\
10  &10 05 09.35 &-7 42 14.5  &25.81  &S  \\
11  &10 05 08.95 &-7 44 37.0  &25.83  &S  \\
12  &10 05 16.44 &-7 41 19.3  &25.84  &S  \\
13  &10 05 18.59 &-7 42 15.2  &25.89  &   \\
14  &10 05 21.25 &-7 40 38.8  &25.93  &S  \\
15  &10 05 06.88 &-7 44 44.0  &25.94  &S  \\
16  &10 05 06.41 &-7 40 47.9  &25.96  &S  \\
17  &10 05 05.04 &-7 45 39.6  &25.97  &S  \\
18  &10 05 11.59 &-7 44 20.2  &25.99  &S  \\
19  &10 05 19.56 &-7 42 25.9  &26.00  &S  \\
20  &10 05 06.28 &-7 41 57.4  &26.03  &S  \\
21  &10 05 09.19 &-7 44 14.9  &26.04  &   \\
22  &10 05 15.94 &-7 40 58.3  &26.05  &S  \\
23  &10 05 21.25 &-7 42 42.1  &26.06  &S  \\
24  &10 05 08.89 &-7 44 24.4  &26.06  &S  \\
25  &10 05 18.78 &-7 42 28.0  &26.07  &S  \\
26  &10 05 07.07 &-7 43 21.5  &26.07  &S  \\
27  &10 05 07.85 &-7 41 01.2  &26.07  &S  \\
28  &10 05 05.41 &-7 43 46.8  &26.08  &S  \\
29  &10 05 08.90 &-7 42 46.1  &26.09  &S  \\
30  &10 05 21.06 &-7 40 05.9  &26.10  &S  \\
31  &10 05 19.13 &-7 42 09.1  &26.11  &S  \\
32  &10 05 19.37 &-7 41 44.9  &26.14  &S  \\
33  &10 05 11.63 &-7 44 28.7  &26.16  &S  \\
34  &10 05 14.37 &-7 41 01.2  &26.16  &S  \\
35  &10 05 20.02 &-7 41 06.4  &26.16  &S  \\
36  &10 05 19.88 &-7 41 32.5  &26.20  &S  \\
37  &10 05 18.03 &-7 42 36.5  &26.20  &   \\
38  &10 05 18.55 &-7 45 21.5  &26.22  &   \\
39  &10 05 11.53 &-7 41 02.6  &26.22  &   \\
40  &10 05 08.50 &-7 43 26.4  &26.27  &   \\
41  &10 05 13.43 &-7 44 46.5  &26.28  &   \\
42  &10 05 05.64 &-7 40 48.8  &26.30  &   \\
43  &10 05 08.92 &-7 44 20.9  &26.31  &   \\
44  &10 05 12.27 &-7 44 37.2  &26.32  &   \\
45  &10 05 05.17 &-7 43 26.4  &26.33  &   \\
46  &10 05 19.24 &-7 42 01.4  &26.35  &   \\
47  &10 05 22.26 &-7 40 28.5  &26.36  &   \\
48  &10 05 07.92 &-7 44 58.1  &26.40  &   \\
49  &10 05 15.99 &-7 40 16.9  &26.42  &   \\
50  &10 05 18.60 &-7 41 53.0  &26.43  &   \\
51  &10 05 11.97 &-7 40 47.2  &26.47  &   \\
52  &10 05 22.59 &-7 45 20.4  &26.53  &   \\
53  &10 05 17.84 &-7 45 00.1  &26.53  &   \\
54  &10 05 05.66 &-7 45 59.1  &26.56  &   \\
55  &10 05 15.85 &-7 44 36.7  &26.63  &   \\
56  &10 05 23.05 &-7 40 14.8  &26.63  &   \\
57  &10 05 18.17 &-7 41 51.7  &26.64  &   \\
58  &10 05 07.36 &-7 45 36.2  &26.66  &   \\
59  &10 05 10.24 &-7 45 55.4  &26.67  &   \\
60  &10 05 17.05 &-7 41 21.6  &26.68  &   \\
61  &10 05 14.16 &-7 44 21.7  &26.94  &   \\
62  &10 05 18.59 &-7 44 37.8  &27.28  &   \\

\enddata
\end{deluxetable}
\clearpage

\begin{deluxetable}{lcccc}
\tablewidth{0pt}
\tablecaption{NGC 3351 Planetaries}
\tablehead{
\colhead{ID} &\colhead{$\alpha(2000)$} &\colhead{$\delta(2000)$}
&\colhead{$m_{5007}$} &\colhead{Sample} \\
}
\startdata

 1  &10 44 00.33  &11 43 30.1  &25.67  &S  \\
 2  &10 43 56.28  &11 44 19.7  &25.71  &S  \\
 3  &10 44 04.80  &11 41 37.7  &25.94  &S  \\
 4  &10 44 01.19  &11 43 21.2  &25.95  &S  \\
 5  &10 43 53.53  &11 41 54.7  &26.00  &S  \\
 6  &10 43 55.88  &11 43 20.5  &26.06  &S  \\
 7  &10 44 06.13  &11 45 32.6  &26.07  &S  \\
 8  &10 43 44.31  &11 43 38.1  &26.14  &S  \\
 9  &10 43 50.72  &11 39 23.9  &26.18  &S  \\
10  &10 43 47.11  &11 44 25.2  &26.21  &S  \\
11  &10 43 53.06  &11 40 38.0  &26.23  &S  \\
12  &10 43 58.18  &11 41 31.7  &26.29  &S  \\
13  &10 44 03.55  &11 40 57.8  &26.35  &   \\
14  &10 43 51.56  &11 39 22.3  &26.35  &   \\
15  &10 43 50.90  &11 43 23.3  &26.37  &   \\
16  &10 44 04.85  &11 40 03.5  &26.40  &   \\
17  &10 43 56.90  &11 39 55.5  &26.46  &   \\
18  &10 43 54.20  &11 40 39.9  &26.48  &   \\
19  &10 43 46.86  &11 41 56.4  &26.51  &   \\
20  &10 44 06.16  &11 40 58.6  &26.61  &   \\

\enddata
\end{deluxetable}
\clearpage

\begin{deluxetable}{lcccc}
\tablewidth{0pt}
\tablecaption{NGC 3627 Planetaries}
\tablehead{
\colhead{ID} &\colhead{$\alpha(2000)$} &\colhead{$\delta(2000)$}
&\colhead{$m_{5007}$} &\colhead{Sample} \\
}
\startdata

 1   &11 20 27.08  &12 58 58.9   &25.65  &S  \\
 2   &11 20 04.26  &13 03 32.3   &25.66  &S  \\
 3   &11 20 25.09  &13 00 35.0   &25.68  &S  \\
 4   &11 20 10.82  &13 01 16.6   &25.70  &S  \\
 5   &11 20 09.80  &13 04 37.9   &25.71  &S  \\
 6   &11 20 10.24  &13 01 23.6   &25.72  &S  \\
 7   &11 20 25.32  &12 58 14.5   &25.75  &S  \\
 8   &11 20 17.49  &12 55 11.9   &25.77  &S  \\
 9   &11 20 08.70  &12 59 42.5   &25.79  &S  \\
10   &11 20 19.23  &13 04 56.7   &25.80  &S  \\
11   &11 20 06.58  &12 56 14.3   &25.82  &S  \\
12   &11 20 25.75  &12 56 41.4   &25.84  &S  \\
13   &11 20 10.92  &12 56 55.1   &25.85  &S  \\
14   &11 20 25.78  &12 57 49.9   &25.86  &S  \\
15   &11 20 10.03  &13 00 40.1   &25.86  &S  \\
16   &11 20 16.81  &13 03 36.7   &25.87  &S  \\
17   &11 20 08.53  &12 57 26.9   &25.88  &S  \\
18   &11 20 06.88  &13 02 40.2   &25.88  &S  \\
19   &11 20 24.43  &12 59 38.8   &25.89  &S  \\
20   &11 20 13.84  &12:55 27.7   &25.89  &S  \\
21   &11 20 09.72  &13 02 42.8   &25.91  &S  \\
22   &11 20 16.73  &13 01 47.5   &25.94  &   \\
23   &11 20 25.92  &12 56 27.6   &25.96  &S  \\
24   &11 20 19.05  &13 01 52.8   &25.98  &S  \\
25   &11 20 16.88  &12 55 14.2   &25.99  &S  \\
26   &11 20 17.01  &13 03 33.7   &26.00  &S  \\
27   &11 20 07.55  &12 57 30.0   &26.01  &S  \\
28   &11 20 14.58  &12 54 43.0   &26.03  &S  \\
29   &11 20 21.14  &12 57 27.8   &26.03  &S  \\
30   &11 20 35.41  &12 59 35.6   &26.04  &S  \\
31   &11 20 04.35  &13 00 27.7   &26.05  &S  \\
32   &11 20 20.65  &13 00 23.4   &26.05  &   \\
33   &11 20 06.15  &13 01 59.7   &26.07  &S  \\
34   &11 20 16.40  &13 02 56.4   &26.07  &S  \\
35   &11 20 18.27  &12 57 06.8   &26.10  &S  \\
36   &11 20 22.19  &12 55 05.3   &26.11  &S  \\
37   &11 20 12.62  &13 04 35.9   &26.13  &S  \\
38   &11 20 06.83  &13 02 52.7   &26.14  &S  \\
39   &11 20 08.10  &12 59 48.4   &26.15  &   \\
40   &11 20 08.35  &13 02 20.0   &26.18  &S  \\
41   &11 19 57.12  &13 03 00.6   &26.18  &S  \\
42   &11 19 51.29  &12 57 13.2   &26.20  &S  \\
43   &11 20 22.39  &12 56 09.4   &26.20  &S  \\
44   &11 20 20.02  &13 05 10.4   &26.22  &   \\
45   &11 20 20.04  &13 04 06.1   &26.22  &   \\
46   &11 20 19.72  &13 02 34.3   &26.22  &   \\
47   &11 20 09.30  &12 57 45.9   &26.23  &   \\
48   &11 20 14.30  &13 02 49.7   &26.27  &   \\
49   &11 20 12.97  &12 56 12.3   &26.29  &   \\
50   &11 20 08.13  &13 04 00.7   &26.30  &   \\
51   &11 20 18.53  &13 02 08.5   &26.31  &   \\
52   &11 20 10.54  &13 02 07.5   &26.32  &   \\
53   &11 20 09.68  &12 57 05.9   &26.36  &   \\
54   &11 19 59.11  &12 58 36.5   &26.37  &   \\
55   &11 20 10.18  &12 56 09.5   &26.40  &   \\
56   &11 20 19.79  &12 55 18.3   &26.41  &   \\
57   &11 20 20.27  &13 04 10.8   &26.42  &   \\
58   &11 20 13.86  &12 56 39.6   &26.43  &   \\
65   &11 20 05.15  &13 04 30.8   &26.45  &   \\
60   &11 20 20.60  &13 01 36.6   &26.45  &   \\
61   &11 20 25.85  &12 58 25.8   &26.45  &   \\
62   &11 20 15.08  &12 56 09.6   &26.50  &   \\
63   &11 20 05.54  &13 05 02.3   &26.53  &   \\
64   &11 20 09.23  &13 04 35.6   &26.54  &   \\
65   &11 20 05.53  &13 02 01.9   &26.63  &   \\
66   &11 20 30.61  &12 55 47.6   &26.66  &   \\
67   &11 20 08.79  &12 55 36.7   &26.68  &   \\
68   &11 19 58.19  &13 03 58.8   &26.69  &   \\
69   &11 20 24.33  &12 59 48.8   &26.70  &   \\
70   &11 20 20.56  &13 00 57.9   &26.73  &   \\
71   &11 20 19.31  &13 02 14.4   &26.73  &   \\
72   &11 20 20.10  &12 54 41.2   &26.77  &   \\
73   &11 20 00.15  &13 00 29.3   &27.04  &   \\
\enddata
\end{deluxetable}

\clearpage

\begin{deluxetable}{lcccc}
\tablewidth{0pt}
\tablecaption{NGC 4258 Planetaries}
\tablehead{
\colhead{ID} &\colhead{$\alpha(2000)$} &\colhead{$\delta(2000)$}
&\colhead{$m_{5007}$} &\colhead{Sample} \\
}
\startdata
1   &12 19 08.60  &47 14 14.0   &25.03  &S   \\
2   &12 18 33.59  &47 20 36.2   &25.06  &S   \\
3   &12 18 46.93  &47 19 45.5   &25.15  &S   \\
4   &12 18 50.73  &47 20 01.4   &25.17  &S   \\
5   &12 18 29.91  &47 21 35.4   &25.20  &S   \\
6   &12 18 45.97  &47 20 15.5   &25.29  &S   \\
7   &12 18 42.40  &47 21 57.2   &25.29  &S   \\
8   &12 18 59.21  &47 20 44.5   &25.30  &S   \\
9   &12 18 47.00  &47 19 47.8   &25.30  &S   \\
10  &12 19 12.17  &47 16 26.9   &25.31  &S   \\
11  &12 18 51.86  &47 21 43.6   &25.33  &S   \\
12  &12 19 04.54  &47 18 46.8   &25.34  &S   \\
13  &12 18 34.94  &47 20 59.1   &25.34  &S   \\
14  &12 18 46.18  &47 21 08.1   &25.36  &S   \\
15  &12 18 33.86  &47 19 36.3   &25.37  &S   \\
16  &12 18 54.99  &47 21 18.2   &25.37  &S   \\
17  &12 18 44.11  &47 22 00.5   &25.37  &S   \\
18  &12 19 14.20  &47 13 46.2   &25.39  &S   \\
19  &12 18 42.41  &47 21 57.3   &25.39  &S   \\
20  &12 18 48.22  &47 17 42.3   &25.40  &S   \\
21  &12 19 15.42  &47 16 51.9   &25.40  &S   \\
22  &12 19 10.77  &47 16 51.5   &25.41  &S   \\
23  &12 19 17.25  &47 15 29.0   &25.43  &S   \\
24  &12 19 13.12  &47 15 57.1   &25.44  &S   \\
25  &12 18 54.81  &47 13 51.7   &25.55  &S   \\
26  &12 19 03.19  &47 13 52.1   &25.56  &S   \\
27  &12 18 29.11  &47 21 47.0   &25.56  &S   \\
28  &12 18 44.42  &47 22 12.9   &25.58  &S   \\
29  &12 18 44.10  &47 21 47.3   &25.60  &   \\
30  &12 18 31.22  &47 20 23.5   &25.60  &   \\
31  &12 19 02.52  &47 14 14.0   &25.62  &   \\
32  &12 19 17.96  &47 14 15.6   &25.62  &   \\
33  &12 19 07.97  &47 16 15.3   &25.63  &   \\
34  &12 19 11.66  &47 16 58.2   &25.65  &   \\
35  &12 19 09.35  &47 14 17.4   &25.66  &   \\
36  &12 18 51.26  &47 16 46.3   &25.66  &   \\
37  &12 18 53.02  &47 16 11.6   &25.66  &   \\
38  &12 19 12.95  &47 20 15.1   &25.68  &   \\
39  &12 18 44.09  &47 19 06.8   &25.68  &   \\
40  &12 18 48.94  &47 21 49.2   &25.69  &   \\
41  &12 19 12.50  &47 17 08.0   &25.70  &   \\
42  &12 18 51.61  &47 20 03.6   &25.72  &   \\
43  &12 18 52.63  &47 15 07.1   &25.73  &   \\
44  &12 19 01.34  &47 20 21.4   &25.74  &   \\
45  &12 18 47.79  &47 19 33.2   &25.75  &   \\
46  &12 18 58.47  &47 15 02.0   &25.75  &   \\
47  &12 18 43.63  &47 21 54.1   &25.77  &   \\
48  &12 18 50.12  &47 15 44.8   &25.80  &   \\
49  &12 18 40.53  &47 16 46.5   &25.85  &   \\
50  &12 19 02.91  &47 15 04.3   &25.86  &   \\
51  &12 19 06.06  &47 19 11.6   &25.87  &   \\
52  &12 18 45.15  &47 19 17.9   &25.88  &   \\
53  &12 18 49.90  &47 22 00.2   &25.91  &   \\
54  &12 18 43.63  &47 19 00.5   &26.01  &   \\
55  &12 19 05.16  &47 20 43.5   &26.03  &   \\
56  &12 18 54.65  &47 21 30.5   &26.04  &   \\
57  &12 18 54.80  &47 20 59.1   &26.14  &   \\
58  &12 19 18.80  &47 16 20.1   &26.16  &   \\

\enddata
\end{deluxetable}

\begin{deluxetable}{lcccc}
\tablewidth{0pt}
\tablecaption{NGC 5866 Planetaries}
\tablehead{
\colhead{ID} &\colhead{$\alpha(2000)$} &\colhead{$\delta(2000)$}
&\colhead{$m_{5007}$} &\colhead{Sample} \\
}
\startdata
 1  &15 06 15.87   &55 46 18.9  &26.35  &S  \\
 2  &15 06 21.38   &55 46 36.5  &26.39  &S  \\
 3  &15 06 17.88   &55 46 09.3  &26.54  &S  \\
 4  &15 06 17.94   &55 47 04.9  &26.56  &S  \\
 5  &15 06 40.23   &55 44 07.4  &26.56  &S  \\
 6  &15 06 27.54   &55 46 46.4  &26.60  &S  \\
 7  &15 06 20.40   &55 46 24.8  &26.64  &S  \\
 8  &15 06 20.19   &55 46 16.8  &26.65  &S  \\
 9  &15 06 22.68   &55 45 48.2  &26.68  &S  \\
10  &15 06 23.10   &55 46 50.2  &26.70  &S  \\
11  &15 06 26.07   &55 45 21.6  &26.89  &   \\
12  &15 06 36.51   &55 44 12.8  &26.93  &   \\
13  &15 06 16.73   &55 44 37.9  &26.99  &   \\
14  &15 06 42.02   &55 44 11.3  &27.05  &   \\
15  &15 06 22.97   &55 45 17.4  &27.07  &   \\
16  &15 06 41.55   &55 46 56.7  &27.11  &   \\
17  &15 06 32.42   &55 44 15.7  &26.42  &   \\
18  &15 06 41.75   &55 45 39.1  &27.45  &   \\
19  &15 06 17.43   &55 44 50.6  &27.60  &   \\
20  &15 06 36.79   &55 44 44.4  &26.72  &   \\
21  &15 06 35.31   &55 43 39.0  &26.74  &   \\
22  &15 06 22.47   &55 45 18.0  &27.76  &   \\
23  &15 06 40.72   &55 45 34.2  &27.88  &   \\
24  &15 06 40.62   &55 45 29.1  &27.89  &   \\

\enddata
\end{deluxetable}

\begin{deluxetable}{lcccccl}
\rotate
\tablewidth{0pt}
\tablecaption{PNLF-Cepheid Comparison}
\tablehead{
\colhead{Galaxy} &\colhead{$E(B-V)$}
&\colhead{Cepheid Distance\tablenotemark{a}} &\colhead{Observed $M^*$}
&\colhead{$12 + \log {\rm O/H}$\tablenotemark{b}}
&\colhead{$\Delta M^*$\tablenotemark{c}} &\colhead{PNLF Reference} \\
}
\startdata
LMC      &0.075 &$18.50$          &$-4.56^{+0.13}_{-0.09}$ &8.50 &0.06
&\citet{p6} \\
SMC      &0.037 &$19.01 \pm 0.03$ &$-4.67^{+0.40}_{-0.17}$ &8.03 &0.48
&\citet{p6} \\
NGC 224  &0.062 &$24.38 \pm 0.05$ &$-4.66^{+0.14}_{-0.11}$ &8.98 &\nodata
&\citet{p2} \\
NGC 300  &0.013 &$26.53 \pm 0.07$ &$-4.21^{+0.67}_{-0.16}$ &8.35 &0.15
&\citet{soffner} \\
NGC 598  &0.041 &$24.56 \pm 0.10$ &$-4.08^{+0.16}_{-0.14}$ &8.82 &\nodata
&\citet{magrini33a}\\
NGC 2403 &0.040 &$27.48 \pm 0.10$ &$-4.41^{+0.16}_{-0.13}$ &8.80 &\nodata
&This paper \\
NGC 3031 &0.080 &$27.75 \pm 0.08$ &$-4.52^{+0.12}_{-0.11}$ &8.75 &\nodata
&\citet{p3} \\
NGC 3351 &0.028 &$29.85 \pm 0.09$ &$-4.39^{+0.19}_{-0.13}$ &9.24 &\nodata
&This paper \\
NGC 3368 &0.025 &$29.97 \pm 0.06$ &$-4.65^{+0.12}_{-0.11}$ &9.20 &\nodata
&\citet{p11} \\
NGC 3627 &0.032 &$29.86 \pm 0.08$ &$-4.44^{+0.12}_{-0.12}$ &9.25 &\nodata
&This paper \\
NGC 4258 &0.016 &$29.44 \pm 0.07$ &$-4.51^{+0.13}_{-0.11}$ &8.85 &\nodata
&This paper \\
NGC 5253 &0.056 &$27.56 \pm 0.14$ &$-4.05^{+0.63}_{-0.16}$ &8.15 &0.33
&\citet{phillips92}\\
NGC 5457 &0.009 &$29.13 \pm 0.11$ &$-4.28^{+0.15}_{-0.14}$ &8.50 &0.06
&\citet{p11} \\

\enddata

\tablenotetext{a}{The SMC Cepheid distance is taken from \citet{udalski99}
assuming $\mu_{\rm LMC} = 18.50$.  All other Cepheid distances are from
\citet{keyfinal} without any correction for metallicity.}
\tablenotetext{b}{Oxygen abundances are taken from \citet{fdatabase}.}
\tablenotetext{c}{Expected metallicity shift in $M^*$ based on the models of
\citet{djv92}.}

\end{deluxetable}

\begin{deluxetable}{lccccl}
\tablewidth{0pt}
\tabletypesize{\footnotesize}
\tablecaption{PNLF-SBF Comparison}
\tablehead{
\colhead{Galaxy} &\colhead{$E(B-V)$} &\colhead{$\mu_{\rm SBF}$}
&\colhead{$\mu_{\rm PNLF}$} &\colhead{$\mu_{\rm PNLF} - \mu_{\rm SBF}$}
&\colhead{PNLF Reference} \\
}
\startdata
NGC 224 &0.062  &$24.36 \pm 0.08$ &$24.36^{+0.09}_{-0.13}$
&$-0.01^{+0.15}_{-0.17}$ &\citet{p2} \\
NGC 891 &0.065  &$29.57 \pm 0.14$ &$29.99^{+0.10}_{-0.13}$
&$+0.40^{+0.19}_{-0.21}$ &\citet{p7} \\
NGC 1023 &0.061 &$30.25 \pm 0.16$ &$29.96^{+0.09}_{-0.10}$
&$-0.29^{+0.20}_{-0.20}$ &\citet{p7} \\
NGC 1316 &0.021 &$31.62 \pm 0.17$ &$31.04^{+0.08}_{-0.09}$
&$-0.59^{+0.19}_{-0.19}$ &\citet{p9} \\
NGC 1399 &0.012 &$31.46 \pm 0.16$ &$31.11^{+0.08}_{-0.09}$
&$-0.36^{+0.18}_{-0.18}$ &\citet{p9} \\
NGC 1404 &0.011 &$31.57 \pm 0.19$ &$31.10^{+0.09}_{-0.12}$
&$-0.48^{+0.21}_{-0.22}$ &\citet{p9} \\
NGC 3031 &0.080 &$27.92 \pm 0.26$ &$27.70^{+0.08}_{-0.09}$
&$-0.22^{+0.29}_{-0.30}$ &\citet{p3} \\
NGC 3115 &0.047 &$29.89 \pm 0.09$ &$30.02^{+0.12}_{-0.15}$
&$+0.12^{+0.16}_{-0.18}$ &This paper \\
NGC 3368 &0.025 &$30.04 \pm 0.22$ &$29.79^{+0.08}_{-0.10}$
&$-0.26^{+0.24}_{-0.24}$ &\citet{p11} \\
NGC 3377 &0.034 &$30.21 \pm 0.09$ &$29.99^{+0.10}_{-0.15}$
&$-0.23^{+0.15}_{-0.18}$ &\citet{p4} \\
NGC 3379 &0.024 &$30.08 \pm 0.11$ &$29.90^{+0.09}_{-0.11}$
&$-0.19^{+0.15}_{-0.16}$ &\citet{p4} \\
NGC 3384 &0.027 &$30.28 \pm 0.14$ &$29.98^{+0.09}_{-0.11}$
&$-0.31^{+0.17}_{-0.18}$ &\citet{p4} \\
NGC 4258 &0.016 &$29.27 \pm 0.14$ &$29.40^{+0.08}_{-0.10}$
&$+0.13^{+0.16}_{-0.17}$ &This paper \\
NGC 4278 &0.030 &$30.99 \pm 0.20$ &$30.00^{+0.09}_{-0.19}$
&$-1.02^{+0.23}_{-0.27}$ &\citet{p10} \\
NGC 4374 &0.042 &$31.28 \pm 0.11$ &$30.89^{+0.09}_{-0.11}$
&$-0.40^{+0.15}_{-0.16}$ &\citet{p5} \\
NGC 4382 &0.030 &$31.29 \pm 0.14$ &$30.73^{+0.08}_{-0.10}$
&$-0.57^{+0.17}_{-0.17}$ &\citet{p5} \\
NGC 4406 &0.029 &$31.13 \pm 0.14$ &$30.92^{+0.08}_{-0.09}$
&$-0.21^{+0.17}_{-0.17}$ &\citet{p5} \\
NGC 4472 &0.022 &$31.02 \pm 0.10$ &$30.69^{+0.10}_{-0.13}$
&$-0.34^{+0.15}_{-0.16}$ &\citet{p5} \\
NGC 4486 &0.023 &$30.99 \pm 0.16$ &$30.71^{+0.08}_{-0.10}$
&$-0.28^{+0.18}_{-0.19}$ &\citet{cm87} \\
NGC 4494 &0.021 &$31.12 \pm 0.11$ &$30.51^{+0.07}_{-0.08}$
&$-0.61^{+0.14}_{-0.14}$ &\citet{p10} \\
NGC 4565 &0.015 &$31.17 \pm 0.17$ &$30.08^{+0.09}_{-0.15}$
&$-1.11^{+0.20}_{-0.22}$ &\citet{p10} \\
NGC 4594 &0.051 &$29.91 \pm 0.18$ &$29.46^{+0.07}_{-0.08}$
&$-0.45^{+0.21}_{-0.21}$ &\citet{ford96} \\
NGC 4649 &0.026 &$31.09 \pm 0.15$ &$30.73^{+0.10}_{-0.13}$
&$-0.37^{+0.18}_{-0.20}$ &\citet{p5} \\
NGC 4697 &0.029 &$30.31 \pm 0.14$ &$29.89^{+0.07}_{-0.07}$
&$-0.42^{+0.16}_{-0.16}$ &\citet{mendez01} \\
NGC 5102 &0.055 &$27.97 \pm 0.13$ &$27.42^{+0.09}_{-0.26}$
&$-0.58^{+0.18}_{-0.28}$ &\citet{mcj} \\
NGC 5128 &0.115 &$28.08 \pm 0.14$ &$27.64^{+0.09}_{-0.09}$
&$-0.44^{+0.23}_{-0.23}$ &\citet{hui93} \\
NGC 5194/95 &0.036 &$29.38 \pm 0.27$ &$29.41^{+0.08}_{-0.12}$
&$+0.01^{+0.29}_{-0.30}$ &\citet{p11} \\
NGC 5866 &0.013 &$30.89 \pm 0.12$ &$30.73^{+0.09}_{-0.12}$
&$-0.17^{+0.15}_{-0.17}$ &This paper \\
\enddata

\end{deluxetable}

\clearpage

\begin{figure}
\vskip-0.4truein
\plotone{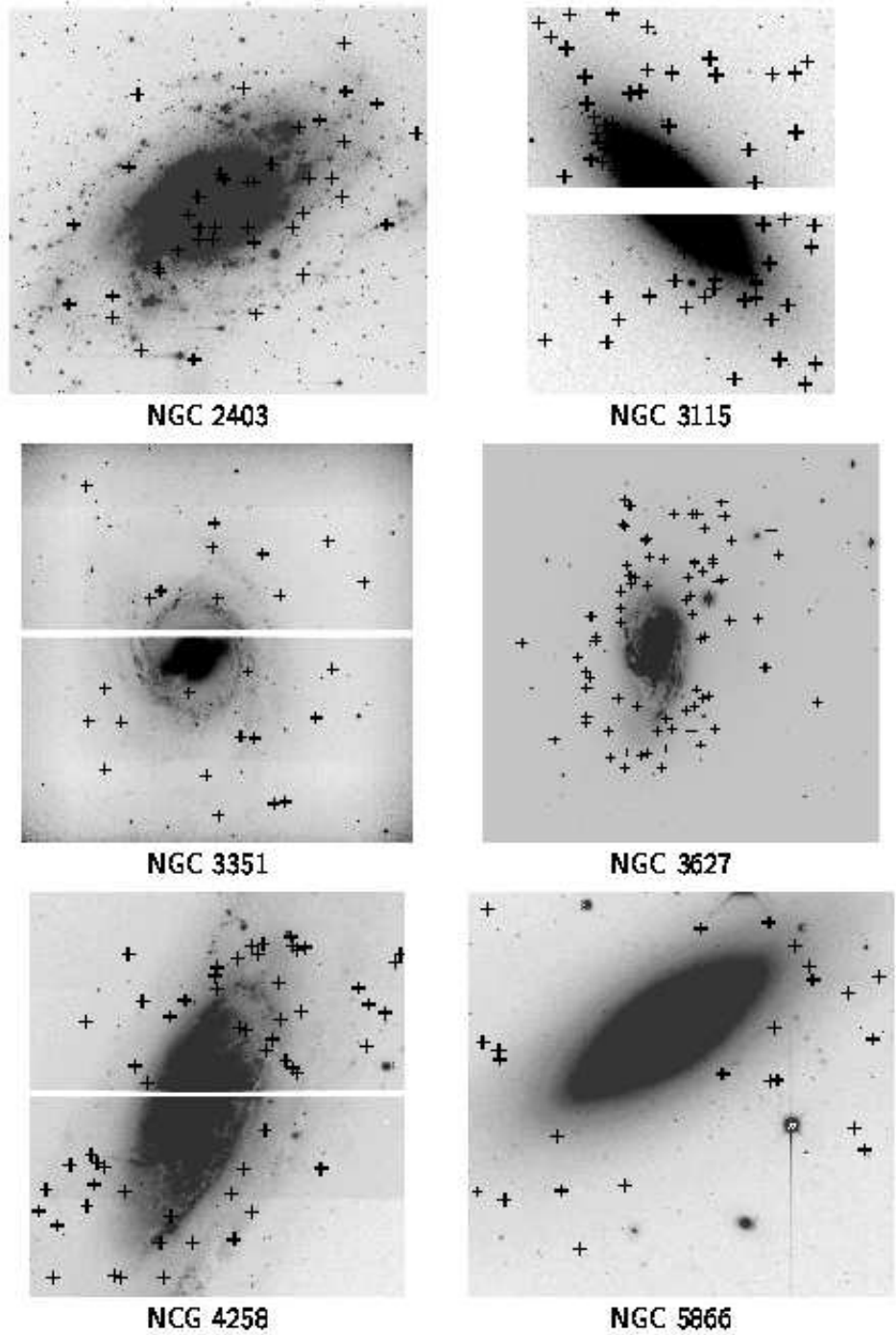} \caption{Our [O~III] $\lambda 5007$ on-band images of the six
galaxies observed in this program.  North is up and east is to the left; the
positions of the PN candidates are marked with crosses.  The fields-of-view are
$16\farcm 4 \times 16\farcm 4$ for NGC~2403 and 3627, $4\farcm 8 \times 4\farcm
8$ for NGC~3351, $6\farcm 4 \times 3\farcm 1$ for NGC~3115, $9\farcm 6 \times
9\farcm 6$ for NGC~4258, and $4\farcm 0 \times 4\farcm 0$ for NGC~5866.
\label{fig1}}
\end{figure}

\begin{figure}
\vskip-0.4truein
\plotone{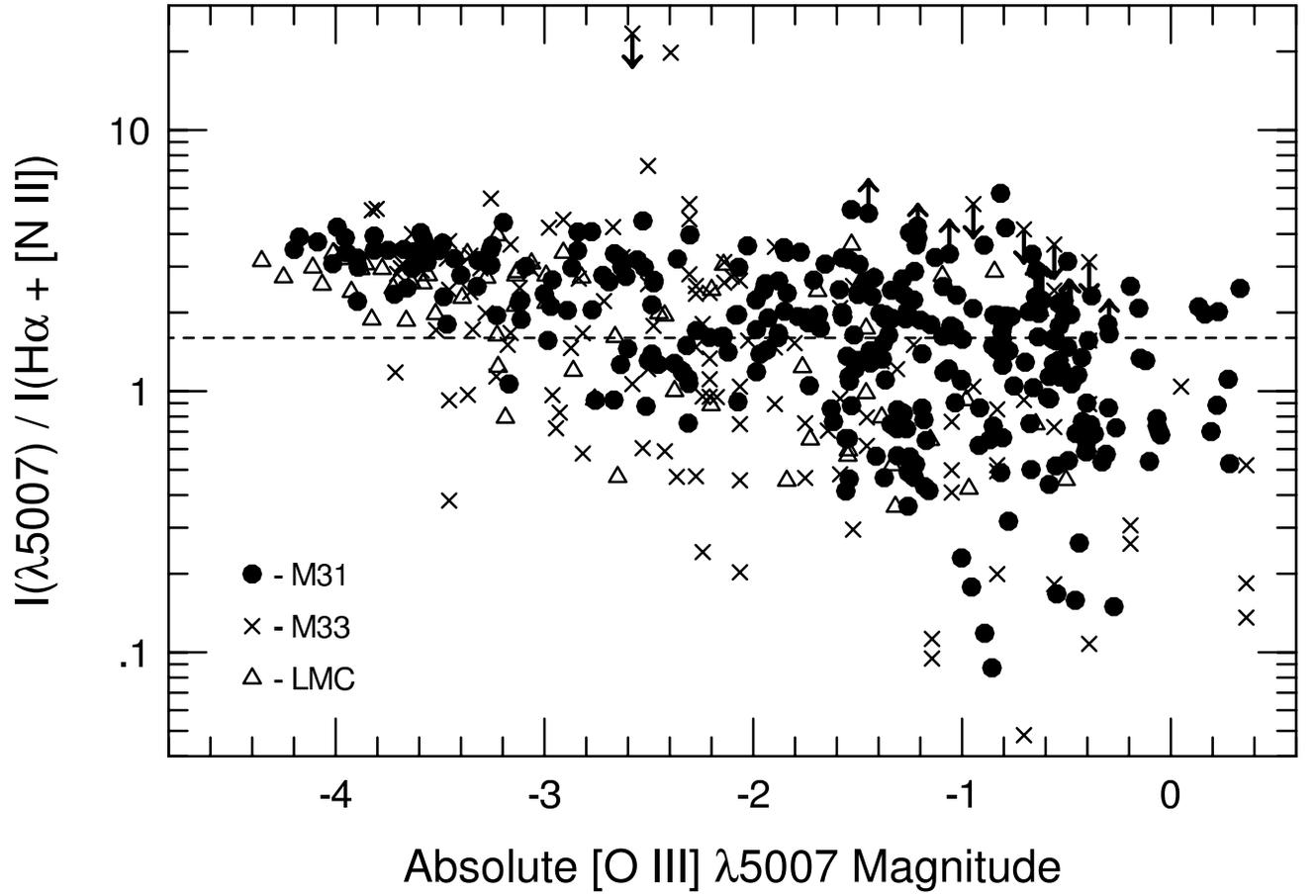}\caption{The [O~III] $\lambda 5007$ to H$\alpha$+[N~II] line
ratios for PNe in the bulge of M31, the disk of M33, and the Large Magellanic
Cloud. The line ratios have been corrected for the effects of foreground
Galactic extinction, but not circumstellar extinction associated with the stars
themselves.  Note that the line ratios of all three stellar populations are
similar, and that PNe in the top $\sim 1$~mag of the PNLF always have [O~III]
$\lambda 5007$ at least twice as bright as H$\alpha$.  The dotted line displays
our line-ratio acceptance criterion for PN candidates. \label{fig2}}
\end{figure}

\begin{figure}
\vskip-0.4truein
\plotone{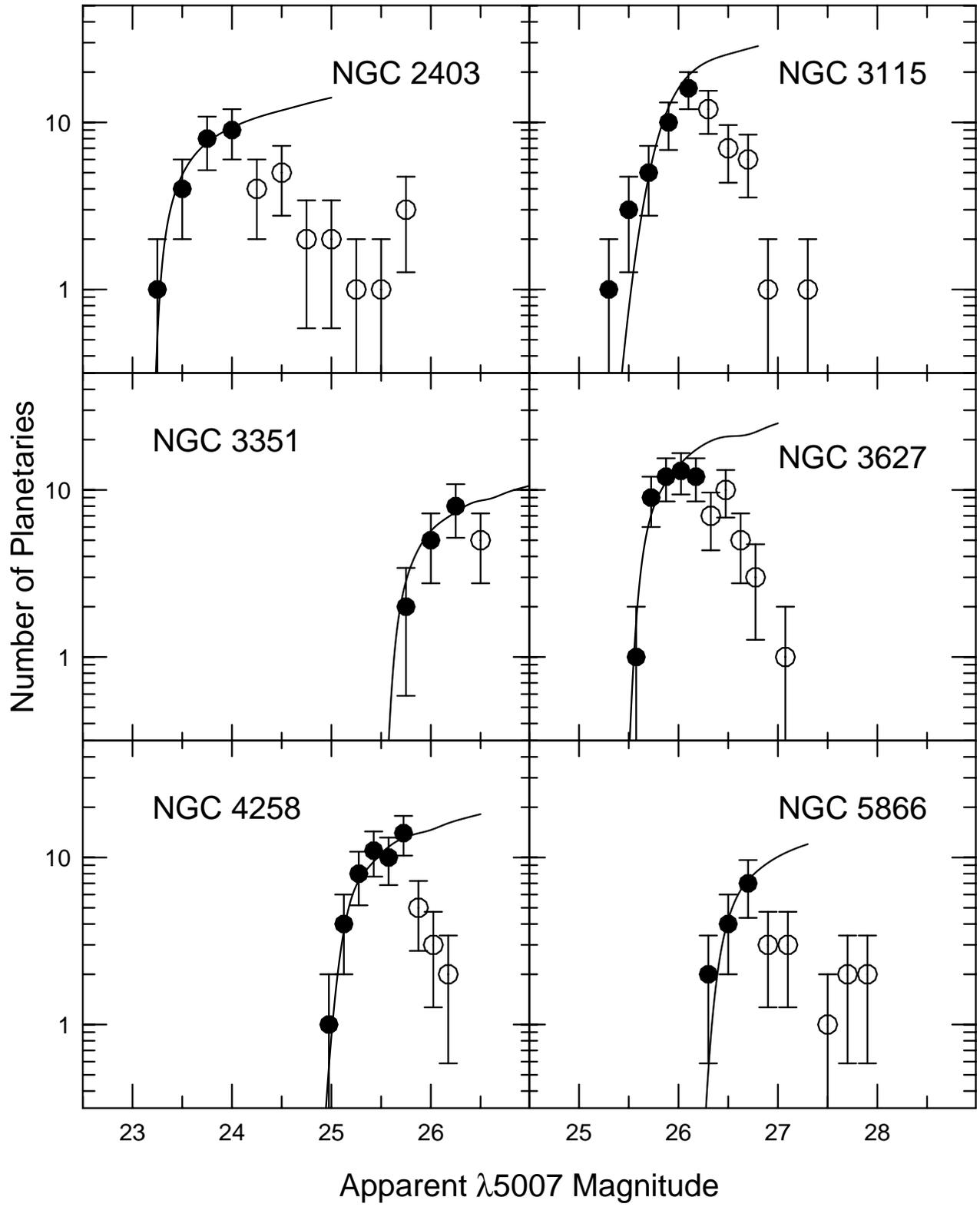}\caption{The observed [O~III] $\lambda 5007$ PNLFs of the six
galaxies studied in this paper.  The curves represent the best fitting
empirical PNLFs convolved with the photometric error function and shifted to
the most likely distance.  The open circles represent points past the
completeness limit. \label{fig3}}
\end{figure}

\begin{figure}
\vskip-0.4truein
\plotone{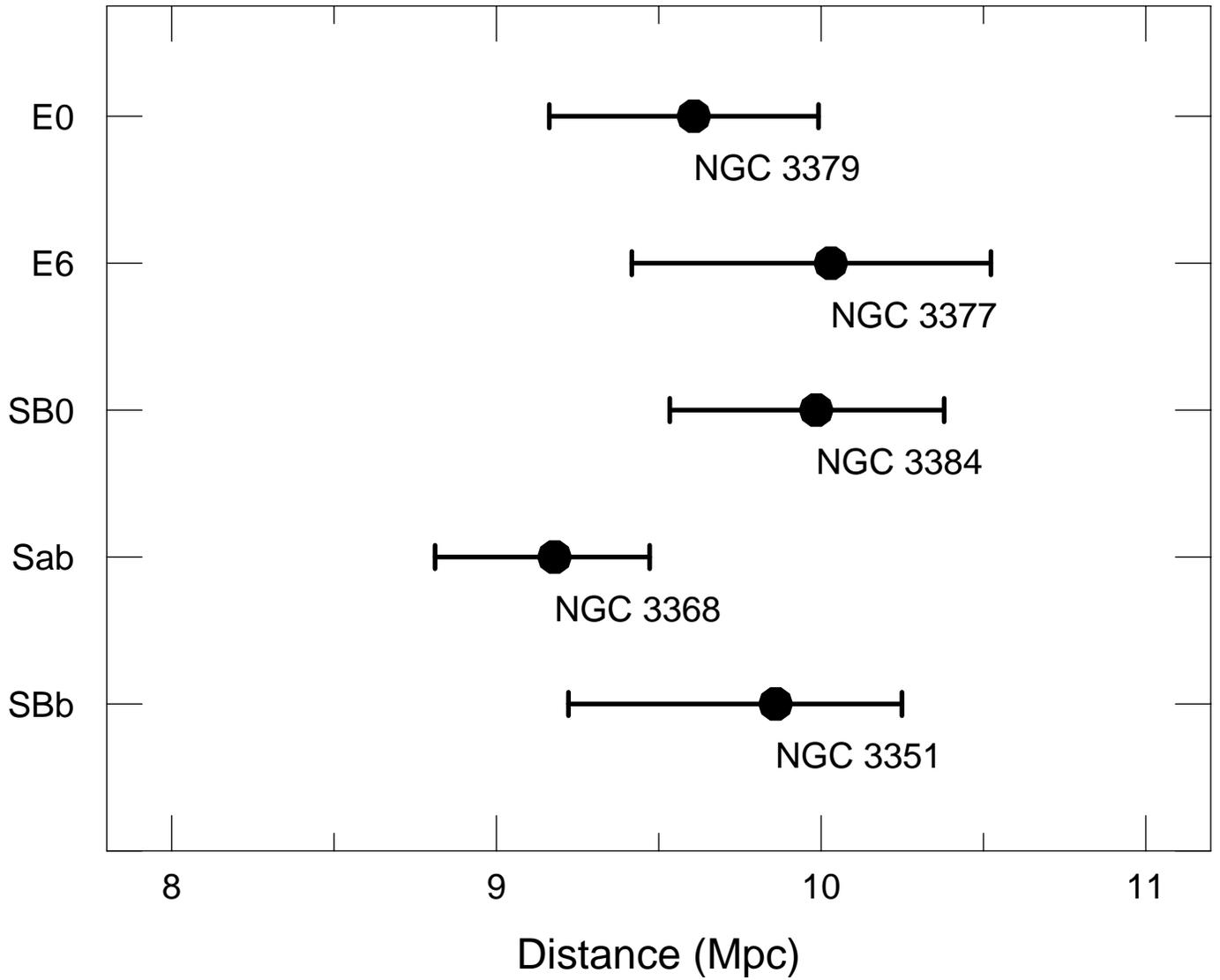}\caption{The PNLF distances to five galaxies of the Leo~I
Group, assuming $M^* = -4.48$.  Note that there is no observable dependence of
PNLF distance on Hubble type:  the technique places all five systems within the
nominal $\sim 1$~Mpc diameter of the group. \label{fig4}}
\end{figure}

\begin{figure}
\vskip-0.4truein
\plotone{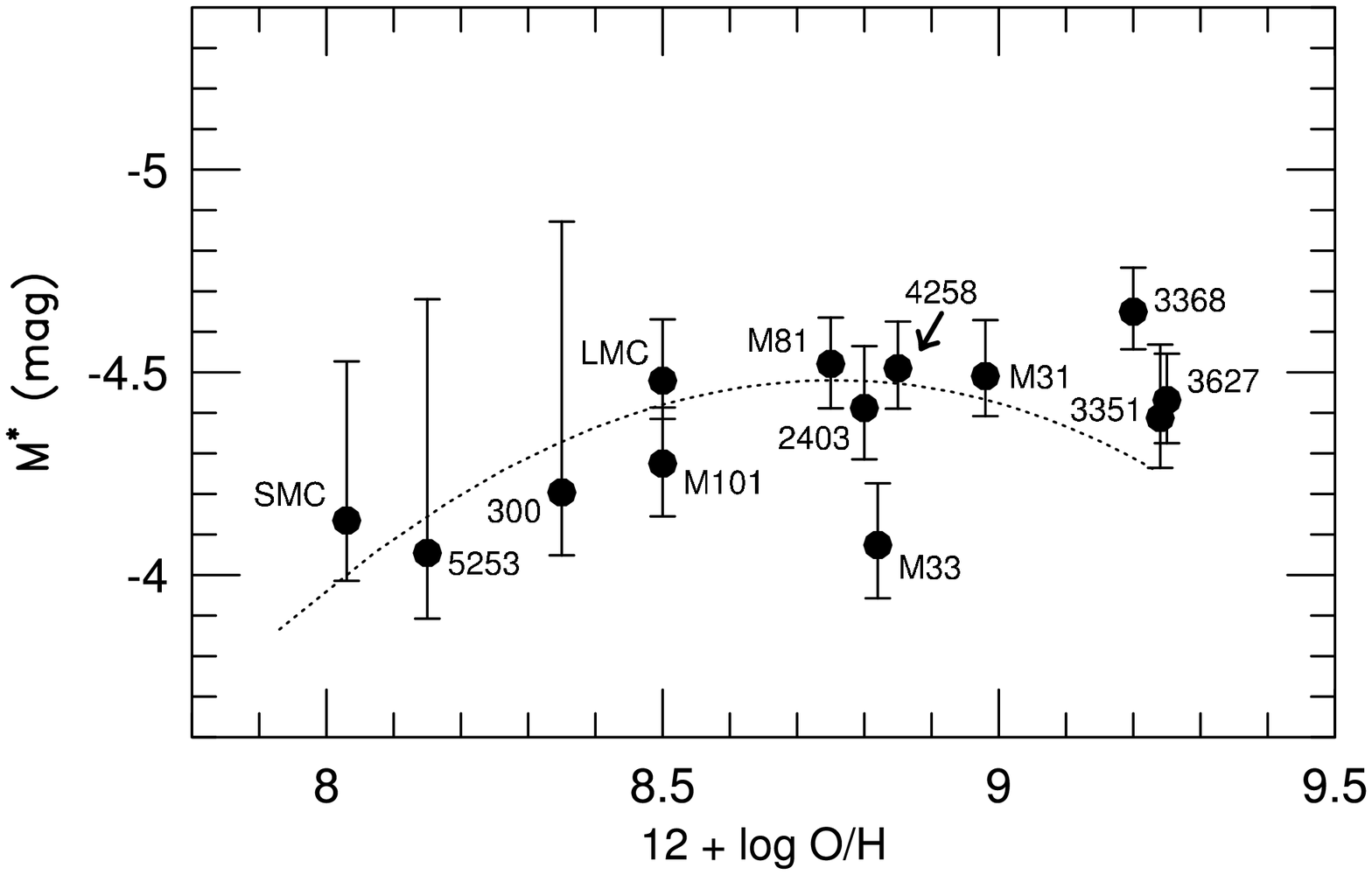}\caption{Values of $M^*$ derived for 13 galaxies using the
Cepheid distances of \citet{keyfinal}, plotted against galactic metallicity, as
determined from the emission lines of H~II regions.  The error bars have been
computed by combining the uncertainties associated with the PNLF fits, the
Cepheid distances, and the Galactic foreground extinction.  The dotted line
shows the \citet{djv92} theoretical dependence of $M^*$ on metallicity. Note
the excellent agreement between the model and the observations on the
low-metallicity side of the curve.  The values of $M^*$ in high-metallicity
galaxies are presumably determined by the galaxies' lower-metallicity stars.
\label{fig5}}
\end{figure}

\begin{figure}
\vskip-0.4truein
\plotone{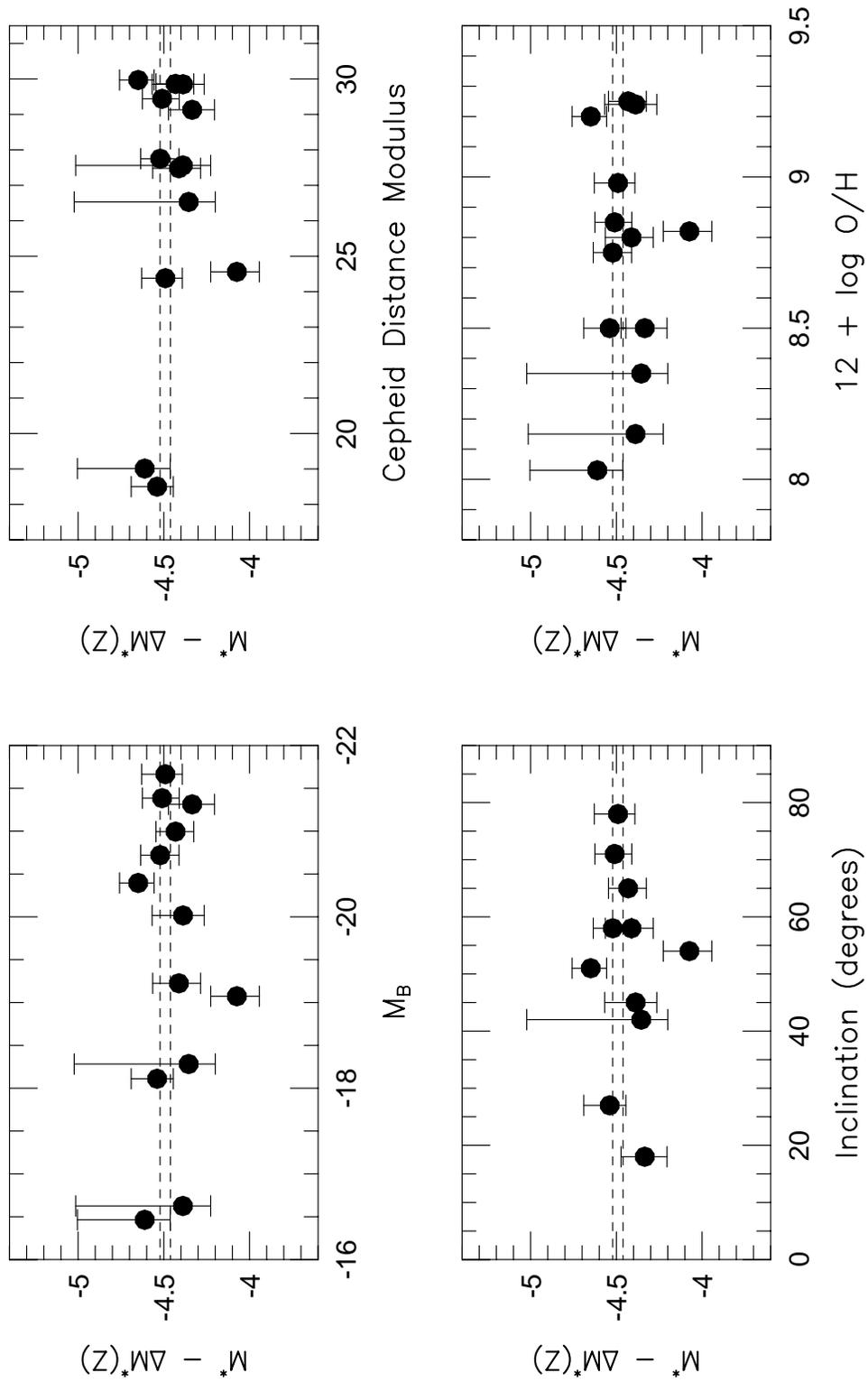}\caption{Metallicity corrected values of $M^*$ derived for 13
galaxies using the Cepheid distances of \citet{keyfinal}, plotted against
galactic absolute magnitude, distance, inclination, and metallicity.  The error
bars have been computed using the uncertainties associated with the PNLF fits,
the Cepheid distances, and the Galactic foreground extinction. The dotted lines
indicate the $1\sigma$ upper and lower limits on the mean value of $M^*$. There
is no evidence for a correlation in any of the plots. \label{fig6}}
\end{figure}

\begin{figure}
\vskip-0.4truein
\plotone{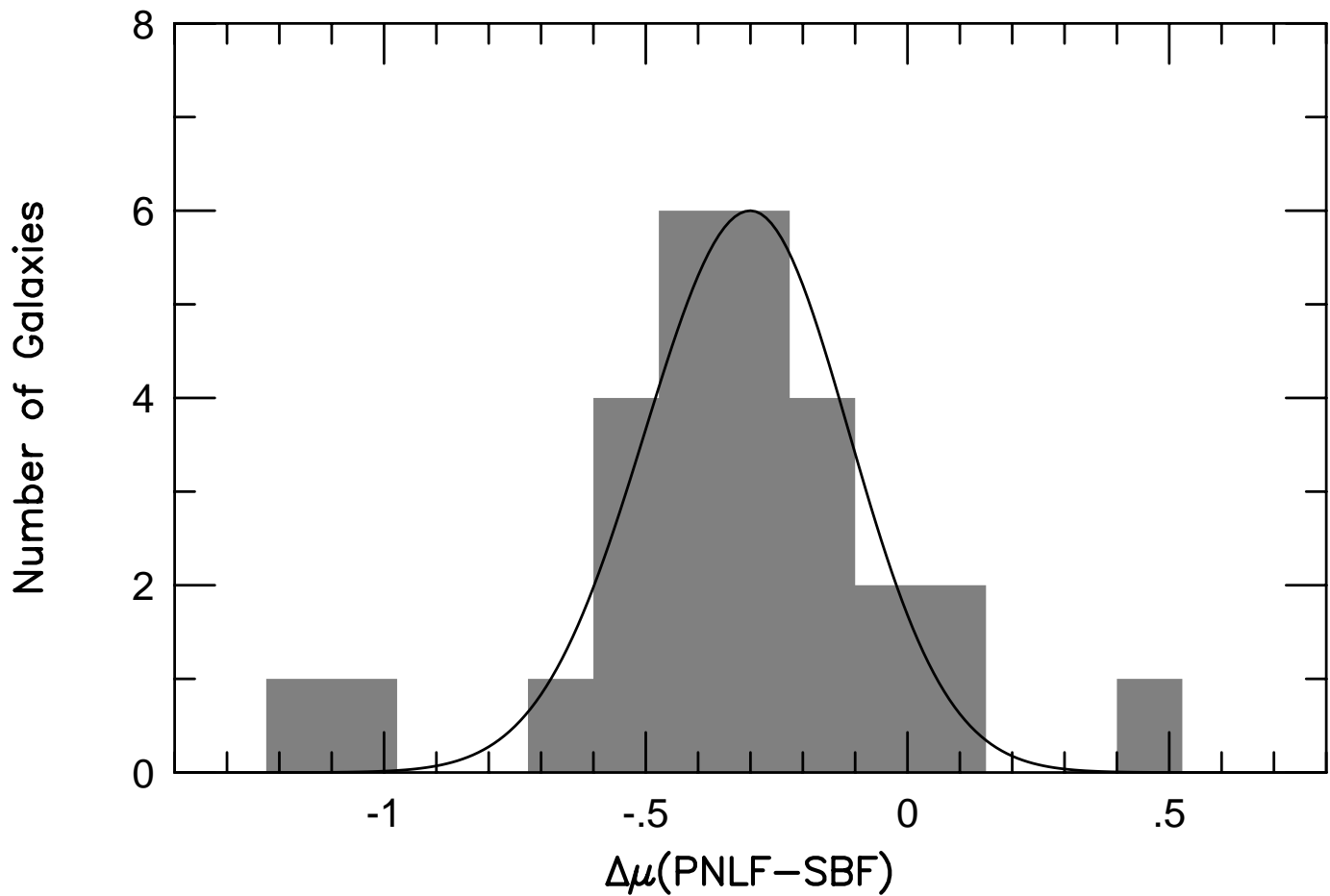}\caption{A histogram of the difference between the PNLF and SBF
distance moduli for 28 galaxies measured by both methods.  The two worst
outliers are the edge-on galaxies NGC~4565 ($\Delta\mu = -0.80$) and NGC~891
($\Delta\mu = +0.71$).  NGC~4258 is also an outlier ($\Delta\mu = -0.70$).  The
curve represents the {\it expected\/} dispersion of the data. The figure
demonstrates that, except for the edge-on galaxies, there is excellent
agreement between the internal and external errors of the methods.
\label{fig7}}
\end{figure}

\begin{figure}
\vskip-0.4truein
\plotone{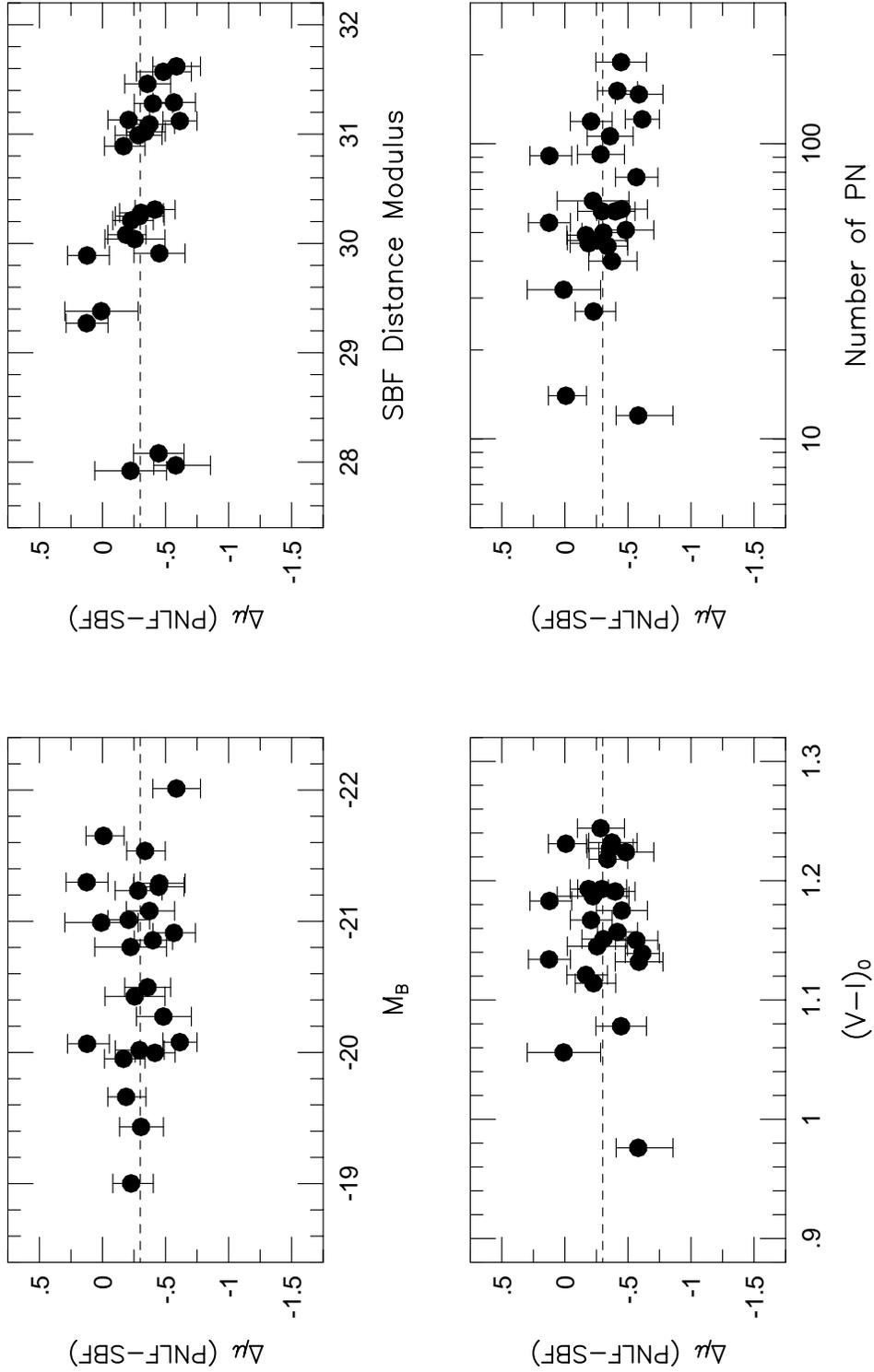}\caption{The difference between SBF and PNLF distance moduli
plotted against galactic absolute magnitude, distance, color, and number of PNe
in the statistical sample.  The three discrepant galaxies, NGC~891, NGC~4565,
and NGC~4278, have not been plotted.  The correlation with SBF distance modulus
is marginally significant ($P \sim 0.1$) due to the low values of the five most
distant objects; if these galaxies are removed from the sample, the
significance of the correlation disappears.  There is no significant
correlation in any of the other panels. \label{fig8}}
\end{figure}

\begin{figure}
\vskip-0.4truein
\plotone{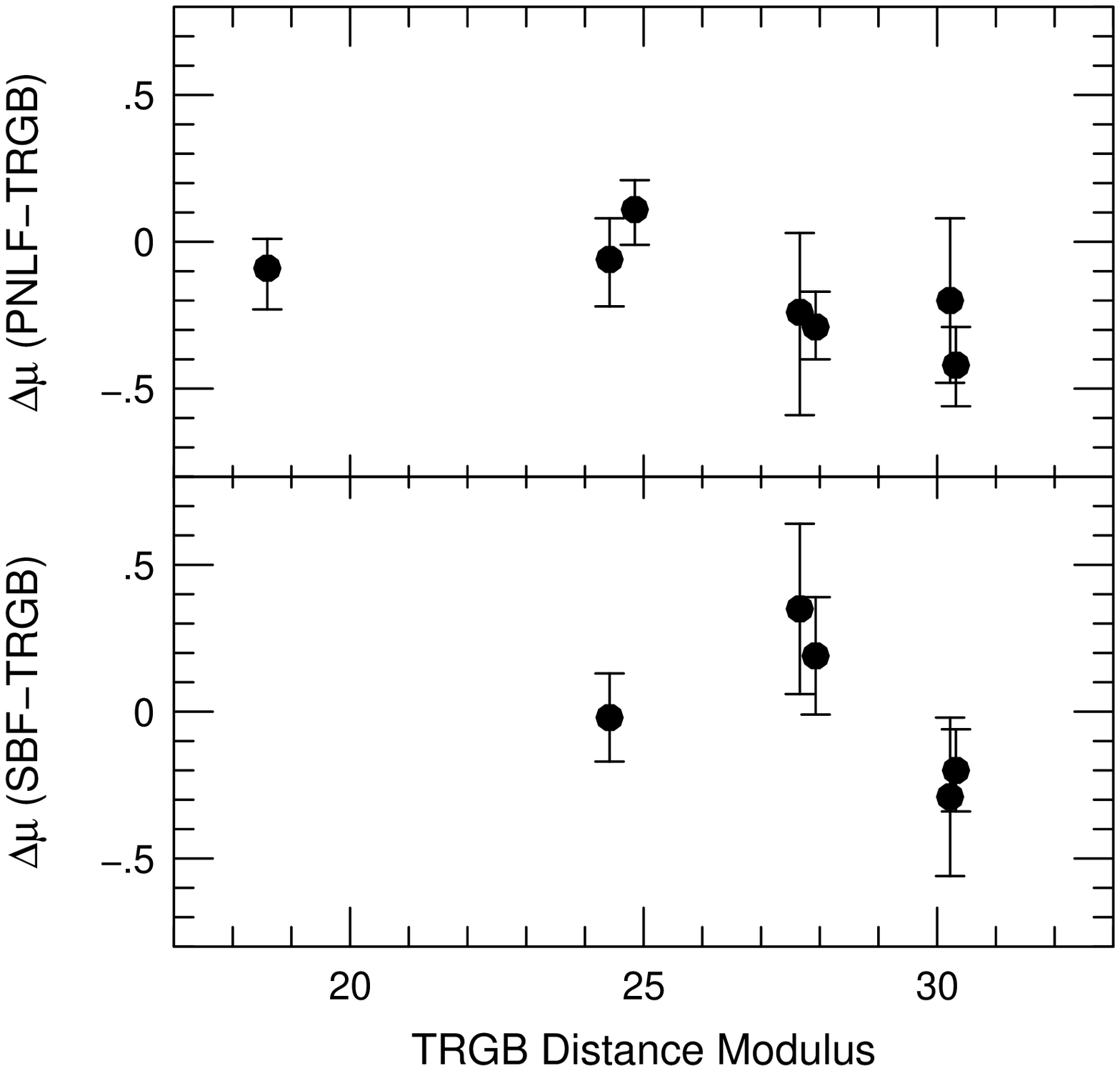}\caption{The upper panel shows the difference between the PNLF
and TRGB distance moduli for galaxies measured with both techniques; the bottom
panel is a similar diagram showing the SBF-TRGB distance residuals. From left
to right, the galaxies displayed are the LMC, M31, M33, NGC~5102, NGC~5128,
NGC~3379, and NGC~3115.  The error bars represent the combined uncertainties of
methods, plus the uncertainty associated with Galactic foreground extinction.
Note the size of the SBF-TRGB error bars in relation to the observed scatter.
This inconsistency, plus the correlation between the PNLF-TRGB distance
residuals and distance, suggests that the TRGB measurements contain an
additional error term. \label{fig9}}
\end{figure}


\clearpage
\begin{thebibliography}{}

\bibitem[Acker \etal(1992)]{acker} Acker, A., Ochsenbein, F., Stenholm, B.,
Tylenda, R., Marcout, J., Schohn, C. 1992, Strasbourg-ESO Catalogue of
Galactic Planetary Nebulae (Garching: ESO)

\bibitem[Arnaboldi \etal(2002)]{arnaboldi02} Arnaboldi, M., Aguerri, J.A.,
Napolitano, N.R., Gerhard, O., Freeman, K.C., Feldmeier, J., Capaccioli, M.,
Kudritzki, R.P., \& M\'endez, R.H. 2002, \apj, 123, 760

\bibitem[Bellazzini, Ferraro, \& Pancino(2001)]{bellazzini} Bellazzini, M.,
Ferraro, F.R., \& Pancino, E. 2001, \apj, 556, 635

\bibitem[Bottinelli \etal(1991)]{bottinelli} Bottinelli, L., Gouguenheim, L.,
Paturel, G., \& Teerikorpi, P. 1991, \aap, 252, 550

\bibitem[Burkhead \& Hutter(1981)]{burk81} Burkhead, M.S., \& Hutter, D.J.
1981, \aj, 86, 523

\bibitem[Burstein \& Heiles(1982)]{bh82} Burstein, D., \& Heiles, C. 1982,
\aj, 87, 1165

\bibitem[Burstein \& Heiles(1984)]{bh84} Burstein, D., \& Heiles, C. 1984,
\apjs, 54, 33

\bibitem[Caputo \etal(2000)]{caputo} Caputo, F., Marconi, M., Musella, I.,
\& Santolamazza, P. 2000, \aap, 359, 1059

\bibitem[Cardelli, Clayton, \& Mathis(1989)]{ccm} Cardelli, J.A.,
Clayton, G.C., \& Mathis, J.S. 1989, \apj, 345, 245

\bibitem[Chromey \etal(1998)]{chromey} Chromey, F.R., Elmegreen, D.M.,
Mandell, A., \& McDermott, J. 1998, \aj, 115, 2331

\bibitem[Ciardullo(1995)]{c95} Ciardullo, R. 1995, in I.A.U. Highlights
of Astronomy 10, ed.~I. Appenzeller (Dordrecht: Kluwer), 507

\bibitem[Ciardullo \etal(2002)]{cfkjg} Ciardullo, R., Feldmeier, J.J.,
Krelove, K., Jacoby, G.H., \& Gronwall, C. 2002, \apj, 566, 784

\bibitem[Ciardullo \etal(1987)]{cfnjs} Ciardullo, R., Ford, H.C., Neill, J.D.,
Jacoby, G.H., and Shafter, A.W. 1987, \apj, 318, 520

\bibitem[Ciardullo \& Jacoby(1992)]{p8} Ciardullo, R., \& Jacoby, G.H. 1992,
\apj, 388, 268

\bibitem[Ciardullo \etal(1998)]{cm87} Ciardullo, R., Jacoby, G.H.,
Feldmeier, J.J., \& Bartlett, R.E. 1998, \apj, 492, 62

\bibitem[Ciardullo \etal(1989)]{p2} Ciardullo, R., Jacoby, G.H., Ford, H.C., \&
Neill, J.D. 1989, \apj, 339, 53

\bibitem[Ciardullo, Jacoby, \& Ford(1989)]{p4} Ciardullo, R., Jacoby, G.H.,
\& Ford, H.C. 1989, \apj, 344, 715

\bibitem[Ciardullo, Jacoby, \& Harris(1991)]{p7} Ciardullo, R., Jacoby, G.H.,
\& Harris, W.E. 1991, \apj, 383, 487

\bibitem[Ciardullo, Jacoby, \& Tonry(1993)]{cjt} Ciardullo, R., Jacoby, G.H.,
\& Tonry, J.L. 1993, \apj, 419, 479

\bibitem[Ciardullo, Kuzio, \& Simone(2001)]{cks} Ciardullo, R., Kuzio, R.E.,
\& Simone, A. 2001, \baas, 33, 1510

\bibitem[Ciatti \& Rosino(1977)]{ciatti} Ciatti, F., \& Rosino, L. 1977,
\aap, 56, 59

\bibitem[de Vaucouleurs(1985)]{deV85} de Vaucouleurs, G. 1985, in ESO Workshop
on the Virgo Cluster of Galaxies, ed.~O.-G. Richter \& B. Binggeli
(Garching: ESO), 413

\bibitem[de Vaucouleurs \& Longo(1988)]{deVLongo} de Vaucouleurs, A., \&
Longo, G. 1988, Catalogue of Visual and Infrared Photometry of Galaxies
from 0.5 Micron to 10 Micron (Austin, University of Texas)

\bibitem[Dopita, Jacoby, \& Vassiliadis(1992)]{djv92} Dopita, M.A.,
Jacoby, G.H., \& Vassiliadis, E. 1992, \apj, 389, 27

\bibitem[Drinkwater, Gregg, \& Colless(2001)]{dgc01}
Drinkwater, M.J., Gregg, M.D., \& Colless, M. 2001, \apj, 548, L139

\bibitem[Durrell, Harris, \& Pritchet(2001)]{durrell} Durrell, P.R., Harris,
W.E., \& Pritchet, C.J. 2001, \aj, 121, 2557

\bibitem[Elson(1997)]{elson} Elson, R.A.W. 1997, \mnras, 286, 771

\bibitem[Feast(1999)]{feast} Feast, M. 1999, \pasp, 111, 775

\bibitem[Feast \& Catchpole(1997)]{fc97} Feast, M.W., \& Catchpole, R.W.
1997, \mnras, 286, L1

\bibitem[Feldmeier, Ciardullo, \& Jacoby(1997)]{p11} Feldmeier, J.J.,
Ciardullo, R., \& Jacoby, G.H. 1997, \apj, 479, 231

\bibitem[Ferrarese \etal(2000a)]{fdatabase} Ferrarese, L., Ford, H.C.,
Huchra, J., Kennicutt, R.C., Mould, J.R., Sakai, S., Freedman, W.L.,
Stetson, P.B., Madore, B.F., Gibson, B.K., Graham, J.A., Hughes, S.M.,
Illingworth, G.D., Kelson, D.D., Macri, L., Sebo, K., \& Silbermann, N.A.
2000a, \apjs, 128, 431

\bibitem[Ferrarese \etal(2000b)]{ferrarese} Ferrarese, L., Mould, J.R.,
Kennicutt, R.C., Huchra, J., Ford, H.C., Freedman, W.L., Stetson, P.B.,
Madore, B.F., Sakai, S., Gibson, B.K., Graham, J.A., Hughes, S.M.,
Illingworth, G.D., Kelson, D.D., Macri, L., Sebo, K., \& Silbermann, N.A.
2000b, \apj, 529, 745

\bibitem[Ford \etal(1996)]{ford96} Ford, H.C., Hui, X., Ciardullo, R.,
Jacoby, G.H., \& Freeman, K.C. 1996, \apj, 458, 455

\bibitem[Freedman \& Madore(1988)]{fm88} Freedman, W.L., \& Madore, B.F.
1988, \apj, 332, L63

\bibitem[Freedman \etal(2001)]{keyfinal} Freedman, W.L., Madore, B.F.,
Gibson, B.K., Ferrarese, L., Kelson, D.D., Sakai, S., Mould, J.R.,
Kennicutt, R.C. Jr., Ford, H.C., Graham, J.A., Huchra, J.P., Hughes,
S.M.G., Illingworth, G.D., Macri, L.M., \& Stetson, P.B. 2001, \apj, 553, 47

\bibitem[Freedman, Wilson, \& Madore(1991)]{freed91} Freedman, W.L.,
Wilson, C.D., \& Madore, B.F. 1991, \apj, 372, 455

\bibitem[Graham \etal(1997)]{graham97} Graham, J.A., Phelps, R.L.,
Freedman, W.L., Saha, A., Ferrarese, L., Stetson, P.B., Madore, B.F.,
Silbermann, N.A., Sakai, S., Kennicutt, R.C., Harding, P., Bresolin,F.,
Turner, A., Mould, J.R., Rawson, D.M., Ford, H.C., Hoessel, J.G., Han, M.,
Huchra, J.P., Macri, L.M., Hughes, S.M., Illingworth, G.D., \& Kelson, D.D.
1997, \apj, 477, 535

\bibitem[Grevesse, Noels, \& Sauval(1996)]{grevesse} Grevesse, N., Noels, A.,
\& Sauval, A.J. 1996, in ASP Conf.~Ser.~99, Cosmic Abundances, eds.~S.S. Holt
\& G. Sonneborn (San Francisco: ASP), 117

\bibitem[Gould \& Uza(1998)]{gould} Gould, A., \& Uza, O. 1998, \apj, 494, 118

\bibitem[Greenhill \etal(1995)]{greenhill} Greenhill, L.J., Henkel, C.,
Becker, R., Wilson, T.L., \& Wouterloot, J.G.A. 1995, \aap, 304, 21

\bibitem[Hamabe \& Okamura(1982)]{hamabe} Hamabe, M., \& Okamura, S.
1982, Ann.~Tokyo Astron.~Obs., 2nd Ser., 18, 191

\bibitem[Harris, Harris, \& Poole(1999)]{harris} Harris, G.L.H., Harris, W.E.,
\& Poole, G.B. 1999, \aj, 117, 855

\bibitem[Herrnstein \etal(1999)] {herrnstein} Herrnstein, J.R., Moran, J.M.,
Greenhill, L.J., Diamond, P.J., Inoue, M., Nakai, N., Miyoshi, M.,
Henkel, C., \& Riess, A. 1999, \nat, 400, 539

\bibitem[Huchra(1987)]{huc87} Huchra, J.P. 1987, in Proc.~13th Texas Symposium
on Relativisitic Astrophysics, ed.~M.P. Ulmer (Singapore: World Scientific), 1

\bibitem[Hui \etal(1993)]{hui93} Hui, X., Ford, H.C., Ciardullo, R., \&
Jacoby, G.H. 1993, \apj, 414, 463

\bibitem[Hui, Ford, \& Jacoby(1994)]{hui94} Hui, X., Ford, H., \& Jacoby, G.
1994, \baas, 26, 938

\bibitem[Humason, Mayall, \& Sandage(1956)]{hms} Humason, M.L., Mayall, N.U.,
\& Sandage, A.R. 1956, \aj, 61, 97

\bibitem[Jacoby \etal(1992)]{mudville} Jacoby, G.H., Branch, D., Ciardullo, R.,
Davies, R.L., Harris, W.E., Pierce, M.J., Pritchet, C.J., Tonry, J.L., \&
Welch, D.L. 1992, \pasp, 104, 599

\bibitem[Jacoby, Ciardullo, \& Ford(1990)]{p5} Jacoby, G.H., Ciardullo, R.,
\& Ford, H.C. 1990, \apj, 356, 332

\bibitem[Jacoby, Ciardullo, \& Harris(1996)]{p10} Jacoby, G.H., Ciardullo, R.,
\& Harris, W.E. 1996, \apj, 462, 1

\bibitem[Jacoby \etal(1989)]{p3} Jacoby, G.H., Ciardullo, R., Ford, H.C., \&
Booth, J. 1989, \apj, 344, 70

\bibitem[Jacoby \& De Marco(2002)]{jd02} Jacoby, G.H., \& De Marco, O. 2002,
\aj, 123, 269

\bibitem[Jacoby, Quigley, \& Africano(1987)]{jqa} Jacoby, G.H., Quigley, R.J.,
\& Africano, J.L. 1987, \pasp, 99, 672

\bibitem[Jacoby, Walker, \& Ciardullo(1990)]{p6} Jacoby, G.H., Walker, A.R.,
\& Ciardullo, R. 1990, \apj, 365, 471

\bibitem[Jensen \etal(2001)]{jensen} Jensen, J.B., Tonry, J.L., Thompson,
R.I., Ajhar, E.A., Lauer, T.R., Rieke, M.J., Postman, M., \& Liu, M.C.
2001, \apj, 550, 503

\bibitem[Karachentsev \etal(2002)]{kara} Karachentsev, I.D., Sharina, M.E.,
Dolphin, A.E., Grebel, E.K., Geisler, D., Guhathakurta, P., Hodge, P.W.,
Karachentseva, V.E., Sarajedini, A., \& Seitzer, P. 2002, \aap, 385, 31

\bibitem[Kelson \etal(2000)]{kelson} Kelson, D.D., Illingworth, G.D.,
Tonry, J.L., Freedman, W.L., Kennicutt, R.C., Mould, J.R., Graham, J.A.,
Huchra, J.P., Macri, L.M., Madore, B.F., Ferrarese, L., Gibson, B.K.,
Sakai, S., Stetson, P.B., Ajhar, E.A., Blakeslee, J.P., Dressler, A.,
Ford, H.C., Hughes, S.M.G., Sebo, K.M., \& Silbermann, N.A. 2000, \apj,
529, 768

\bibitem[Kennicutt \etal(1998)]{kenn98} Kennicutt, R.C., Stetson, P.B.,
Saha, A., Kelson, D., Rawson, D.M., Sakai, S., Madore, B.F., Mould, J.R.,
Freedman, W.L., Bresolin, F., Ferrarese, L., Ford, H., Gibson, B.K.,
Graham, J.A., Han, M., Harding, P., Hoessel, J.G., Huchra, J.P.,
Hughes, S.M.G., Illingworth, G.D., Macri, L.M., Phelps, R.L., Silbermann, N.A.,
Turner, A.M., \& Wood, P.R. 1998, \apj, 498, 181

\bibitem[Kim \etal(2002)]{kim} Kim, M., Kim, E., Lee, M.G., Sarajedini, A.,
\& Geisler, D. 2002, \aj, 123, 244

\bibitem[Kochanek(1997)]{kochanek} Kochanek, C.S. 1997, \apj, 491, 13

\bibitem[Kormendy \& Bahcall(1974)]{kb74} Kormendy, J., \& Bahcall, J.N.
1974, \aj, 79, 671

\bibitem[Kraan-Korteweg, Cameron, \& Tammann(1988)]{kkct88}Kraan-Korteweg, R.C.,
Cameron, L.M., \& Tammann, G.A. 1988, \apj, 331, 620

\bibitem[Layden \etal(1996)]{layden} Layden, A.C., Hanson, R.B., Hawley, S.L.,
Klemola, A.R., \& Hanley, C.J. 1996, \aj, 112, 2110

\bibitem[Lee, Freedman \& Madore(1993)]{lee93} Lee, M.G., Freedman, W.L.,
\& Madore, B.F. 1993, \apj, 417, 553

\bibitem[Macri \etal(2000)]{macri} Macri, L.M., Huchra, J.P., Sakai, S.,
Mould, J.R., \& Hughes, S.M. 2000, \apjs, 128, 461

\bibitem[Madore \& Freedman(1995)]{mf95} Madore, B.F., \& Freedman, W.L. 1995,
\aj, 109, 1645

\bibitem[Magrini \etal(2000)]{magrini33a} Magrini, L., Corradi, R.L.M.,
Mampaso, A., \& Perinotto, M. 2000, \aap, 355, 713

\bibitem[Magrini \etal(2001a)]{magrini33b} Magrini, L., Cardwell, A.,
Corradi, R.L.M., Mampaso, A., \& Perinotto, M. 2001a, \aap, 367, 498

\bibitem[Magrini \etal(2001b)]{magrini81} Magrini, L., Perinotto, M.,
Corradi, R.L.M., \& Mampaso, A. 2001b, \aap, 379, 90

\bibitem[Massey \etal(1988)]{massey} Massey, P., Strobel, K., Barnes, J.V., \&
Anderson, E. 1988, \apj, 328, 315

\bibitem[McMillan, Ciardullo, \& Jacoby(1993)]{p9} McMillan, R.,
Ciardullo, R., \& Jacoby, G.H. 1993, \apj, 416, 62

\bibitem[McMillan, Ciardullo, \& Jacoby(1994)]{mcj} McMillan, R.,
Ciardullo, R., \& Jacoby, G.H. 1994, \aj, 108, 1610

\bibitem[M\'endez \etal(1993)]{mendez93} M\'endez, R.H., Kudritzki, R.P.,
Ciardullo, R., \& Jacoby, G.H. 1993, \aap, 275, 534

\bibitem[M\'endez \etal(2001)]{mendez01} M\'endez, R.H., Riffeser, A.,
Kudritzki, R.-P., Matthias, M., Freeman, K.C., Arnaboldi, M.,
Capaccioli, M., \& Gerhard, O.E. 2001, \apj, 563, 135

\bibitem[McClure \& Racine(1969)]{mcclure} McClure, R.D., \& Racine, R. 1969,
\aj, 74, 1000

\bibitem[Meatheringham \& Dopita(1991a)]{md1} Meatheringham, S.J., \&
Dopita, M.A. 1991a, \apjs, 75, 407

\bibitem[Meatheringham \& Dopita(1991b)]{md2} Meatheringham, S.J., \&
Dopita, M.A. 1991b, \apjs, 76, 1085

\bibitem[Miyoshi \etal(1995)]{miyoshi} Miyoshi, M., Moran, J., Herrnstein,
J., Greenhill, L., Nakai, N., Diamond, P., \& Inoue, M. 1995, \nat, 373, 127

\bibitem[Monet \etal(1998)]{monet} Monet, D., Bird, A., Canzian, B.,
Dahn, C., Guetter, H., Harris, H., Henden, A., Levine, S., Luginbuhl, C.,
Monet, A.K.B., Rhodes, A., Riepe, B., Sell, S., Stone, R., Vrba, F.,
\& Walker, R. 1998, PMM USNO-A2.0: A Catalogue of Astrometric Standards
(Washington, DC: US Naval Obs.)

\bibitem[Newman \etal(2001)]{newman01} Newman, J.A., Ferrarese, L.,
Stetson, P.B., Maoz, E., Zepf, S.E., Davis, M., Freedman, W.L., \&
Madore, B.F. 2001, \apj, 553, 562

\bibitem[Panagia(1999)]{panagia99} Panagia, N. 1999, in IAU Symp.~190,
ed.~Chu, Y.-H., Suntzeff, N., Hesser, J., \& Bohlender, D.
(Dordrecht, Kluwer), 549

\bibitem[Panagia \etal(1991)]{panagia91} Panagia, N., Gilmozzi, R.,
Macchetto, F., Adorf, H.-M., \& Kirshner, R.P. 1991, \apj, 380, L23

\bibitem[Peimbert \& Torres-Peimbert(1981)]{peimbert2} Peimbert, M., \&
Torres-Peimbert, S. 1981, \apj, 245, 845

\bibitem[Persson, Frogel, \& Aaronson(1979)]{persson} Persson, S.E.,
Frogel, J.A., \& Aaronson, M. 1979, \apjs, 39, 61

\bibitem[Phillips \etal(1992)]{phillips92} Phillips, M.M., Jacoby, G.H.,
Walker, A.R., Tonry, J.L., \& Ciardullo, R. 1992, \baas, 24, 749

\bibitem[Pottasch(1990)]{pottasch90} Pottasch, S.R. 1990, \aap, 236, 231

\bibitem[Reid(1997)]{reid} Reid, I.N. 1997, \aj, 114, 161

\bibitem[Richer(1993)]{richer} Richer, M.G. 1993, \apj, 415, 240

\bibitem[Rots(1978)]{rots} Rots, A.H. 1978, \aj, 83, 219

\bibitem[Lasker \etal(1990)]{gsc} Lasker, B.M., Sturch, C.R., McLean, B.J.,
Russell, J.L., Jenkner, H., \& Shara, M.M. 1990, \aj, 99, 2019

\bibitem[Saha \etal(1999)]{saha99} Saha, A, Sandage, A., Tammann, G.A.,
Labhardt, L., Macchetto, F.D., \& Panagia, N. 1999, \apj, 522, 802

\bibitem[Sakai \etal(1997)]{sakai} Sakai, S., Madore, B.F., Freeman, W.L.,
Lauer, T.R., Ajhar, E.A., \& Baum, W.A. 1997, \apj, 478, 49

\bibitem[Sakai, Zaritsky, \& Kennicutt(2000)]{szk} Sakai, S., Zaritsky, D.,
\& Kennicutt, R.C. 2000, \aj, 119, 1197

\bibitem[Sandage \& Tammann(1982)]{st82}Sandage, A., \& Tammann, G.A. 1982,
\apj, 256, 339

\bibitem[Sasselov \etal(1997)]{sasselov} Sasselov, D.D., Beaulieu, J.P.,
Renault, C., Grison, P., Ferlet, R., Vidal-Madjar, A., Maurice, E.,
Pr\'evot, L., Aubourg, E., Bareyre, P., Brehin, S., Coutures, C.,
Delabrouille, N., de Kat, J., Gros, M., Laurent, B., Lachi\`eze-Rey, M.,
Lesquoy, E., Magneville, C., Milsztajn, A., Moscoso, L., Queinnec, F.,
Rich, J., Spiro, M., Vigroux, L., Zylberajch, S., Ansari, R., Cavalier, F.,
Moniez, M., Gry, C., Guibert, J., Moreau, O., \& Tajhmady, F. 1997, \aap, 324,
471.

\bibitem[Schlegel, Finkbeiner, \& Davis(1998)]{schlegel} Schlegel, D.J.,
Finkbeiner, D.P., \& Davis, M. 1998, \apj, 500, 525

\bibitem[Shaver \etal(1983)]{shaver} Shaver, P.A., McGee, R.X., Newton, L.M.,
Danks, A.C., \& Pottasch, S.R. 1983, \mnras, 204, 53

\bibitem[Soffner \etal(1996)]{soffner} Soffner, T., M\'endez, R.H.,
Jacoby, G.H., Ciardullo, R., Roth, M.M., \& Kudritzki, R.P. 1996, \aap, 306, 9

\bibitem[Stetson(1987)]{stet87}Stetson, P.B. 1987, \pasp, 99, 191

\bibitem[Stetson(1992)]{stet92}Stetson, P.B. 1992, in ASP Conf.~Ser.~25,
Astronomical Data Analysis Software and Systems I., eds. D.M. Worral,
C. Biemesderfer, \& J. Barnes (San Francisco: ASP), 297

\bibitem[Stetson, Davis, \& Crabtree(1990)]{stet90} Stetson, P.B., Davis, L.E.,
\& Crabtree, D.R. 1990, in ASP~Conf.~Ser.~8, CCDs in Astronomy,
ed.~G.H. Jacoby (San Francisco: ASP), 289.

\bibitem[Stone(1977)]{stone} Stone, R.P.S. 1977, \apj, 218, 767

\bibitem[Tammann \& Sandage(1968)]{ts68} Tammann, G.A., \& Sandage, A. 1968,
\apj, 151, 825

\bibitem[Tonry \etal(2001)]{tonry} Tonry, J.L., Dressler, A., Blakeslee, J.P.,
Ajhar, E.A., Fletcher, A.B., Luppino, G.A., Metzger, M.R., \& Moore, C.B.
2001, \apj, 546, 681

\bibitem[Udalski(2000)]{udalski00} Udalski, A. 2000, \apj, 531, L25

\bibitem[Udalski \etal(1999)]{udalski99} Udalski, A., Szyma\'nski, M.,
Kubiak, M., Pietrzy\'nski, G., Soszy\'nski, I., Wo\'zniak, P., \&
\.Zebru\'n, K. 1999, Act.~Astr., 49, 201

\bibitem[Udalski \etal(2001)]{udalski01} Udalski, A., Wyrzykowski, L.,
Pietrzy\'nski, G., Szewczyk, O., Szyma\'nski, M., Kubiak, M.,
Soszy\'nski, I., \& \.Zebru\'n, K. 2001, Act.~Astr., 51, 221

\bibitem[Van Dyk \etal(2000)]{vandyk} Van Dyk, S.D., Peng, C.Y., King, J.Y.,
Filippenko, A.V., Treffers, R.R., Li, W., \& Richmond, M.W. 2000,
\pasp, 112, 1532

\bibitem[Vassiliadis \etal(1992)]{vdm} Vassiliadis, E., Dopita, M.A., Morgan,
D.H., \& Bell, J.F. 1992, \apjs, 83, 87

\bibitem[Watson \& Wallin(1994)]{ww94} Watson, W.D., \& Wallin, B.K. 1994,
\apj, 432, L35

\bibitem[Watanabe \etal(2001)]{wat01} Watanabe, M., Yasuda, N., Itoh, N.,
Ichikawa, T., \& Yanagisawa, K. 2001, \apj, 555, 215

\bibitem[Welch \etal(1986)]{welch} Welch, D.L., McAlary, C.W., McLaren, R.A.,
\& Madore, B.F. 1986, \apj, 305, 583

\bibitem[Wells \etal(1994)]{wells} Wells, L.A., Phillips, M.M., Suntzeff, N.B.,
Heathcote, S.R., Hamuy, M., Navarrete, M., Fernandez, M., Weller, W.G.,
Schommer, R.A., Kirshner, R.P., Leibundgut, B., Willner, S.P., Peletier, R.F.,
Schlegel, E.M., Wheeler, J.C., Harkness, R.P., Bell, D.J., Matthews, J.M.,
Filippenko, A.V., Shields, J.C., Richmond, M.W., Jewitt, D., Luu, J.,
Tran, H.D., Appleton, P.N., Robson, E.I., Tyson, J.A., Guhathakurta, P.,
Eder, J.A., Bond, H.E., Potter, M., Veilleux, S., Porter, A.C.,
Humphreys, R.M., Janes, K.A., Williams, T.B., Costa, E., Ru\'iz, M.T.,
Lee, J.T., Lutz, J.H., Rich, R.M., Winkler, P.F., \& Tyson, N.D. 1994,
\aj, 108, 2233

\bibitem[West \& Blakeslee(2000)]{wb00} West, M.J., \& Blakeslee, J.P.
2000, \apj, 543, L27

\end{thebibliography}
\end{document}